\documentclass[conference, 10pt]{IEEEtran}
\ifCLASSINFOpdf
\else
\fi
\usepackage{graphicx}
\usepackage{color}
\usepackage{placeins}
\usepackage{float}
\usepackage{tabularx,colortbl}
\graphicspath{{figures/Appendix/}}
\usepackage{amssymb} 
\usepackage{amsmath} 
\usepackage[breaklinks=true]{hyperref}
\usepackage{setspace}
\usepackage{cite}
\usepackage{subfigure}
\def \Cfi {C_{f}}
\def \Csi {C_{s}}

\begin{document}

\title{Denoising Using Projection Onto Convex Sets (POCS) Based Framework}

\author{\IEEEauthorblockN { Mohammad Tofighi, Kivanc Kose$^*$, A. Enis Cetin}
\IEEEauthorblockA{Dept. of Electrical and Electronic Engineering, Bilkent University,  Ankara, Turkey\\
$^*$Dermatology Department, Memorial Sloan Kettering Cancer Center, New York, New York\\
tofighi@ee.bilkent.edu.tr, $^*$kosek@mskcc.org, cetin@bilkent.edu.tr\\}
}

\maketitle

\begin{abstract}
Two new optimization techniques based on projections onto convex space (POCS) framework for solving convex optimization problems are presented. The dimension of the minimization problem is lifted by one and  sets corresponding to the cost function are defined. If the cost function is a convex function in $R^N$ the corresponding set is also a convex set in $R^{N+1}$. The iterative optimization approach starts with an arbitrary initial estimate in $R^{N+1}$ and an orthogonal projection is performed onto one of the sets in a sequential manner at each step of the optimization problem. The method provides globally optimal solutions in total-variation (TV), filtered variation (FV), $\ell_1$, and entropic cost functions. A new denoising algorithm using the TV framework is developed. The new algorithm does not require any of the regularization parameter adjustment. Simulation examples are presented.\end{abstract}


\IEEEpeerreviewmaketitle

\section{Introduction}
\label{sec:Introduction}
In many inverse signal and image processing problems and compressing sensing problems an optimization problem is solved to find a solution to the following problem:
\begin{equation}
\label{app:eq:c1}
\underset{\mathbf{w}\in \text{C}}{\text{min}} f(\mathbf{w}),
\end{equation}
where $\mathbb{C}$ is a set in $\mathbb{R}^N$ and $f(\mathbf{w})$ is the cost function.
Some commonly used cost functions are based on $\ell_1$, $l_2$, total-variation (TV), filtered variation, and entropic functions \cite{Rud92, Bar07,Can08, Kos12,Gunay}.
%
Bregman developed iterative methods based on the so-called Bregman distance  to solve convex optimization problems which arise in signal and image processing \cite{Bre67}. In Bregman's approach, it is necessary to perform a D-projection (or Bregman projection) at each step of the algorithm and it may not be easy to compute the Bregman distance in general \cite{Yin08,Kiv12,Gunay}.

In this article Bregman's older projections onto convex sets (POCS) framework \cite{Bregman,You82} is used to solve convex and some non-convex optimization problems instead of the Bregman distance approach. Bregman's POCS method has also been widely used for finding a common point of convex sets in many inverse signal and image processing problems\cite{You82,Her95,Cen12,Sla08,Cet03,Cetin94,Cetin89,Kose11,Cen81,Sla09,The11,censor1987optimization,Tru85,Com04,Com93,Kim92,yamada2011minimizing,censor1987some,Sez82,censor1992proximal,Tuy81,censor1981row,censor1991optimization,Ros13}. In the ordinary POCS approach the goal is simply to find a vector which is in the intersection of convex sets.  In each step of the iterative algorithm an orthogonal projection is performed onto one of the convex sets. Bregman showed that successive orthogonal projections converge to a vector which is in the intersection of all the convex sets. If the sets do not intersect iterates oscillate between members of the sets \cite{Gub67,Com12,Cet97}. Since  there is no need to compute the Bregman distance in standard POCS, it found applications in many practical problems.

In our approach the dimension of the minimization problem is lifted by one and  sets corresponding to the cost function are defined. This approach is graphically illustrated in Fig. \ref{app:convex}. If the cost function is a convex function in $R^N$ the corresponding set is also a convex set in $R^{N+1}$. As a result the convex minimization problem is reduced to finding a specific member (the optimal solution) of the set corresponding to the cost function. As in ordinary POCS approach the new iterative optimization method starts with an arbitrary initial estimate in $R^{N+1}$ and an orthogonal projection is performed onto one of the sets. After this vector is calculated it is projected onto the other set. This process is continued in a sequential manner at each step of the optimization problem.
The method provides globally optimal solutions in total-variation, filtered variation, $\ell_1$, and entropic function based cost functions because they are convex cost functions.

The article is organized as follows. In Section \ref{sec:Convex Minimization}, the convex minimization method based on the POCS approach is introduced. In Section \ref{sec:Denoising using POCS}, a new denoising method based on the convex minimization approach introduced in Section \ref{sec:Convex Minimization}, is presented.This new approach uses supporting hyperplanes of the TV function and it does not require a regularization parameter as in other TV based methods. Since it is very easy to perform an orthogonal projection onto a hyperplane this method is computationally implementable for many cost functions without solving any nonlinear equations. In Section~\ref{sec:Simulation Results}, we present the simulation results and some denoising examples.

\section{Convex Minimization}
\label{sec:Convex Minimization}
Let us first consider a convex minimization problem
\begin{equation}
\label{app:eq:c5}
\underset{\mathbf{w}\in \mathbb{R}^N}{\text{min}} f(\mathbf{w}),
\end{equation}
where $f:\mathbb{R}^N \rightarrow \mathbb{R}$ is a convex function.
We increase the dimension by one to define the following sets in $\mathbb{R}^{N+1}$ corresponding to the cost function $f(\mathbf{w})$ as follows:
\begin{equation}
\label{app:eq:c6}
\text{C}_f = \{\underline{\mathbf{w}} = [\mathbf{w}^T~y]^T : \mathrm{~} y\geq f(\mathbf{w})\},
\end{equation}
which is the set of $N+1$ dimensional vectors whose $(N+1)^{st}$ component $y$ is greater than $f(\mathbf{w})$. This set $C_{f}$ is called the epigraph of $f$. We use bold face letters for $N$ dimensional vectors and underlined bold face letters for $N+1$ dimensional vectors, respectively.

\begin{figure}[ht!]
\begin{center}
\noindent
\includegraphics[width=80mm]{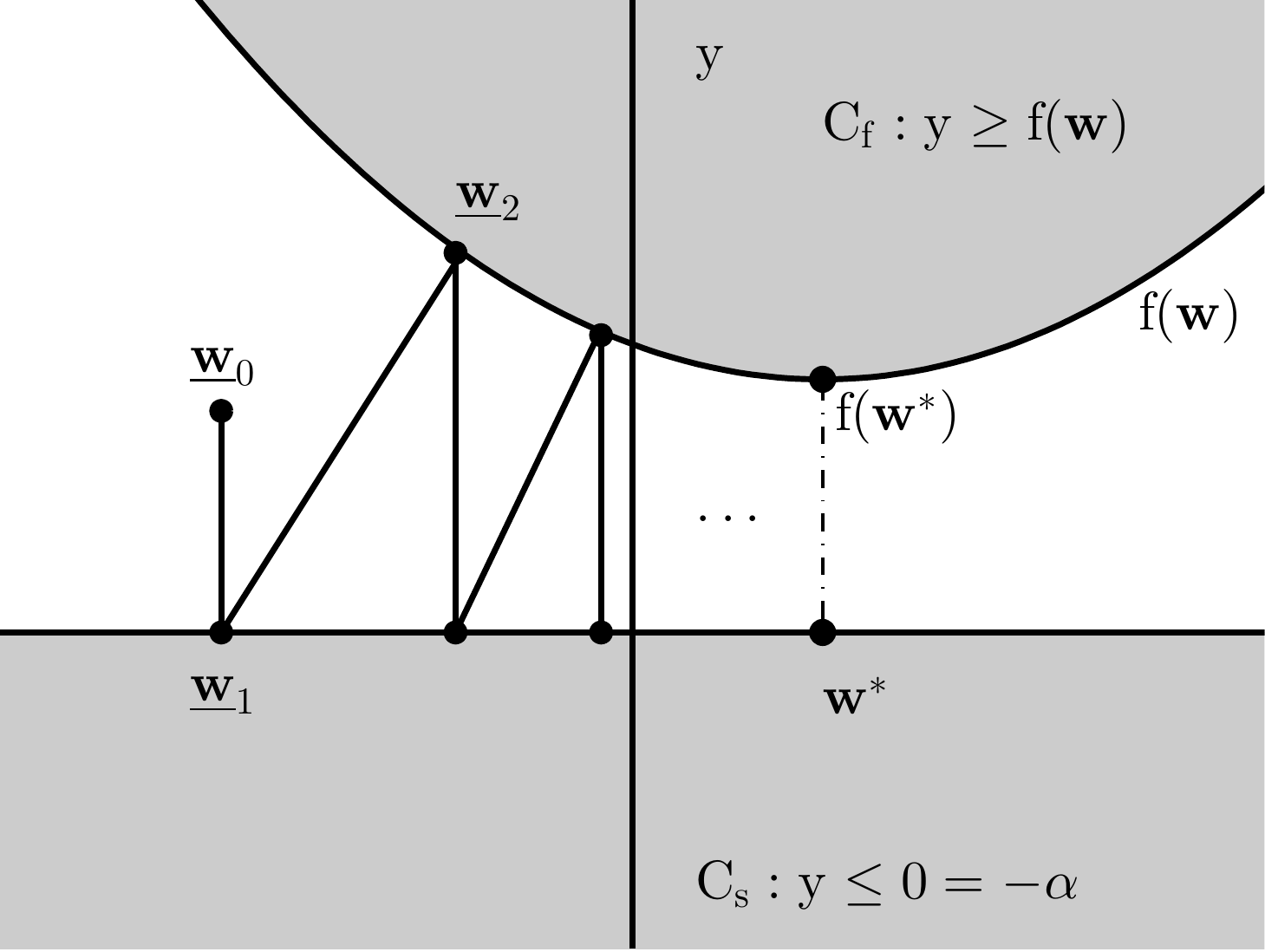}
\caption[Two projecting convex sets.]{Two convex sets $\Cfi$ and $\Csi$ corresponding to the cost function $f$. We sequentially project an initial vector $\underline{\mathbf{w}}_0$ onto $\Csi$ and C$\Cfi$ to find the global minimum which is located at $\mathbf{w}^*$.}
\label{app:convex}
\end{center}
\end{figure}

The second set that is related with the cost function $f(\mathbf{w})$ is the level set:
\begin{equation}
\label{app:eq:c7}
\text{C}_s = \{\underline{\mathbf{w}} = [\mathbf{w}^T~y]^T : ~ y\leq \alpha , ~\underline{\mathbf{w}} \in \mathbb{R}^{N+1}\},
\end{equation}
where $\alpha$ is a real number. Here it is assumed that $f(\mathbf{w})\geq \alpha$ for all $f(\mathbf{w})\in \mathbb{R}$ such that the sets $\Cfi$ and $\Csi$ do not intersect. They are both closed and convex sets in  $\mathbb{R}^{N+1}$. Sets $\Cfi$ and $\Csi$ are graphically illustrated in Fig. \ref{app:convex} in which $\alpha=0.$

The POCS based minimization algorithm starts with an arbitrary $\underline{\mathbf{w}}_0 =[ \mathbf{w}_0^T ~ y_0]^T \in \mathbb{R}^{N+1}$. We project $\underline{\mathbf{w}}_0$ onto the set $\Csi$ to obtain the first iterate $\underline{\mathbf{w}}_1$ which will be,
\begin{equation}
\underline{\mathbf{w}}_1 = [\mathbf{w}_0^T~0]^T,
\end{equation}
where $\alpha=0$ is assumed as in Fig. \ref{app:convex}. Then we project $\underline{\mathbf{w}}_1$ onto the set $\Cfi$. The new iterate  $\underline{\mathbf{w}}_2$ is determined by minimizing the distance between $\underline{\mathbf{w}}_1$ and $\Cfi$, i.e.,

\begin{equation}
\label{app:eq:convex}
\underline{\mathbf{w}}_2 = \text{arg} \underset{\underline{\mathbf{w}}\in \text{C}_{\mathrm{s}}}{\text{min}} \|\underline{\mathbf{w}}_1 - \underline{\mathbf{w}}\|.
\end{equation}

Equation \ref{app:eq:convex} is the ordinary orthogonal projection operation onto the set $\mathrm{C_f} \in \mathbb{R}^{N+1}$.
To solve the problem in Eq. \ref{app:eq:convex} we do not need to compute the Bregman's so-called D-projection. After finding $\underline{\mathbf{w}}_2$, we perform the next projection onto the set $\Csi$ and obtain $\underline{\mathbf{w}}_3$ etc. Eventually iterates oscillate between two nearest vectors of the two sets $\Csi$ and $\Cfi$. As a result we obtain
\begin{equation}
\label{app:eq:convex2}
\underset{n \rightarrow \infty}{\text{lim}} \underline{\mathbf{w}}_{2n} = [\mathbf{w}^*~f(\mathbf{w}^*)]^T,
\end{equation}
where $\mathbf{w}^*$ is the N dimensional vector minimizing $f(\mathbf{w})$. The proof of Eq. \ref{app:eq:convex2} follows from Bregman's POCS theorem \cite{Bregman,Gub67}. It was generalized to non-intersection case by Gubin et. al \cite{Gub67,Cen12},\cite{Com12}. Since the two closed and convex sets $\Csi$ and $\Cfi$ are closest to each other at the optimal solution case, iterations oscillate between the vectors $[\mathbf{w}^*~f(\mathbf{w}^*)]^T$ and $[\mathbf{w}^*~0]^T$ in $R^{N+1}$ as $n$ tends to infinity. It is possible to increase the speed of convergence by non-orthogonal projections \cite{Com93}.

If the cost function $f$ is not convex and have more than one local minimum then the corresponding set $\Cfi$ is not convex in $R^{N+1}$. In this case iterates may converge to one of the local minima.

\section{Denoising Using POCS}
\label{sec:Denoising using POCS}
\begin{figure}[ht!]
\begin{center}
\noindent
\includegraphics[width=90mm]{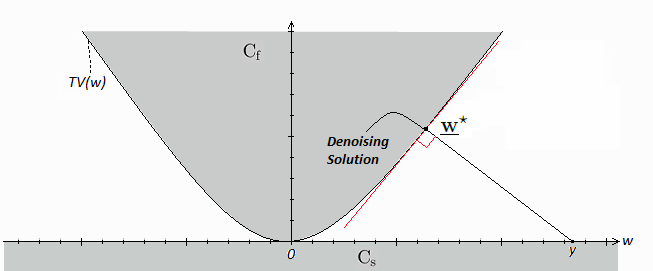}
\caption{Graphical representation of the minimization of (\ref{app:eq:6}). y is projected onto the set $\Cfi$. TV(\textbf{w}) is zero for $\textbf{w}=[0, 0,...,0]^{T}$ or when it is a constant vector.}
\label{app:convexSol}
\end{center}
\end{figure}

In this section, we present a new method of denoising, based on TV and FV. Let the noisy signal be \textbf{y}, and the original signal or image be $\textbf{w}_{0}$. Suppose that the observation model is the additive noise model:

\begin{equation}
\textbf{y} = \textbf{w}_{0} + \textbf{v},
\end{equation}
where \textbf{v} is the additional noise. In this approach we solve the following problem for denoising:

\begin{equation}
\label{app:eq:6}
\underline{\mathbf{w}}^{\star} = \text{arg} \underset{\underline{\mathbf{w}}\in \text{C}_{\mathrm{f}}}{\text{min}} \|\underline{\mathbf{y}} - \underline{\mathbf{w}}\|^{2},
\end{equation}
where, $\underline{\mathbf{y}}$ = [$y^{T}$\ 0] and $\Cfi$ is the epigraph set of TV or FV in $R^{N+1}$. The minimization problem is essentially the orthogonal projection onto the set $\Cfi$. This means that we select the nearest vector $\underline{\mathbf{w}}^{\star}$ on the set $\Cfi$ to \textbf{y}. This is graphically illustrated in Fig. \ref{app:convexSol}.

\begin{figure}[ht!]
\begin{center}
\noindent
\includegraphics[width=90mm]{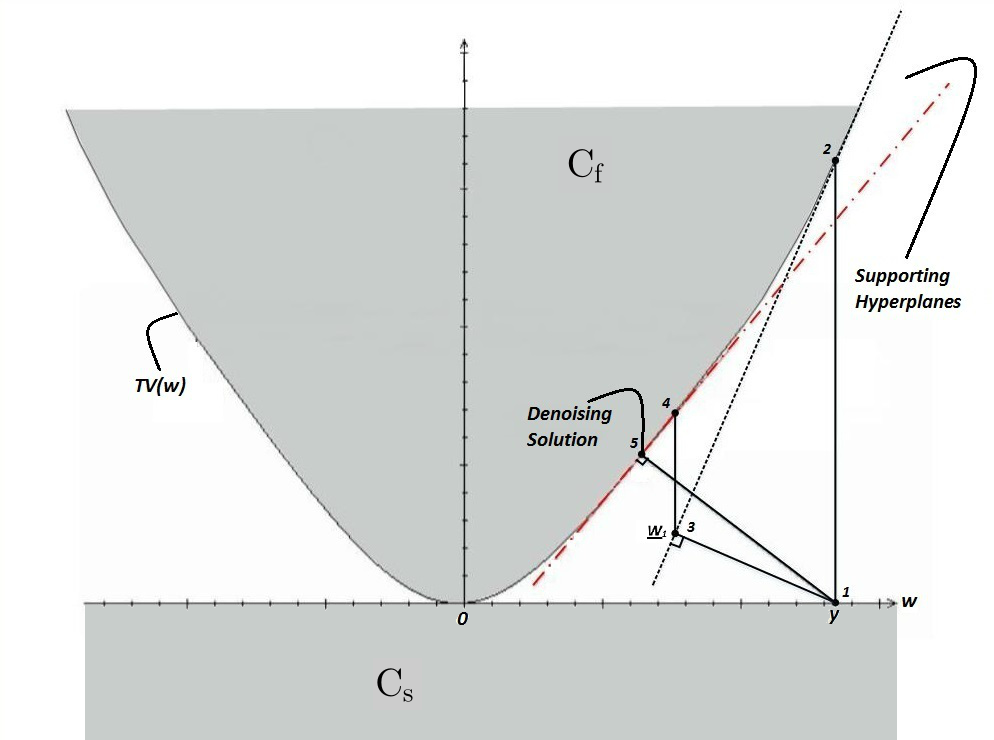}
\caption{Graphical representation of the minimization of (\ref{app:eq:6}), using projection onto the supporting hyperplanes of $\Cfi$.}
\label{app:convex1}
\end{center}
\end{figure}

In current TV based denoising methods \cite{Chambolle, Com11} the following cost function is used:

\begin{equation}
\label{app:eq:cost}
{{\text{min}} \|\underline{\mathbf{y}} - \underline{\mathbf{w}}\|}^2_2 + \lambda \text{TV}(\textbf{w}).
\end{equation}
The solution of this problem can be obtained using the method that we discussed in Section \ref{sec:Convex Minimization}. One problem with this approach is the estimation of the regularization parameter $\lambda$. One has to determine the $\lambda$ in an ad hoc manner or by visual inspection. On the other hand we do not require any parameter adjustment in (\ref{app:eq:6}).

The denoising solution in (\ref{app:eq:6}) can be found by performing successive orthogonal projection onto supporting hyperplanes of the epigraph set $\Cfi$. In the first step we calculated TV($\mathbf{y}$). We also calculate the surface normal at $\underline{\mathbf{y}}$ = [$\mathbf{y}^{T}$ \ TV($\mathbf{y}$)] in $R^{N+1}$ and determine the equation of the supporting hyperplane at [$\mathbf{y}^{T}$ \ TV($\mathbf{y}$)].
We project $\underline{\mathbf{y}}$ = [$\mathbf{y}^{T}$ \ 0] onto this hyperplane and obtain $\underline{\mathbf{w}}_1$ as our first estimate as shown in Fig. \ref{app:convex1}. In the second step we project $\underline{\mathbf{w}}_1$ onto the set $\Csi$ by simply making its last component zero. We calculate the TV of this vector and the surface normal, and the supporting hyperplane as in the previous step. We project $\underline{\mathbf{y}}$ onto the new supporting hyperplane, etc.

\begin{figure}[ht!]
\begin{center}
\noindent
\includegraphics[width=90mm]{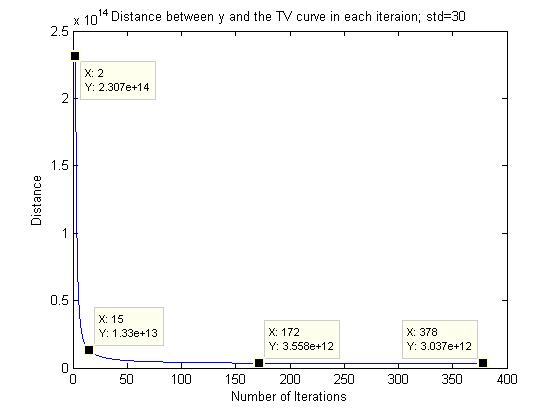}
\caption{Orthogonal distance from y to the epigraph of TV in each iteration.}
\label{app:graphdist}
\end{center}
\end{figure}

The sequence of iterations obtained in this manner converges to a vector in the intersection of $\Csi$ and $\Cfi$. In this problem the sets $\Csi$ and $\Cfi$ intersect because $TV(\textbf{w})=0$ for $\textbf{w}=[0, 0,...,0]^{T}$ or for a constant vector. However, we do not want to find a trivial constant vector in the intersection of $\Csi$ and $\Cfi$. We calculate the distance between $\underline{\mathbf{y}}$ and $\underline{\mathbf{w}}_i$ at each step of the iterative algorithm described in the previous paragraph. This distance ${ \|\underline{\mathbf{y}} - \underline{\mathbf{w}}\|}^2_2$ initially decreases and starts increasing as $i$ increases. Once we detect the increase we perform some refinement projections to obtain the solution of the denoising problem. A typical convergence graph is shown in Fig. \ref{app:graphdist} for the ``note" image.
Simulation exapmles are presented in the next section.

\section{Simulation Results}
\label{sec:Simulation Results}
\begin{figure}[ht!]
\begin{center}
\noindent
\includegraphics[width=70mm]{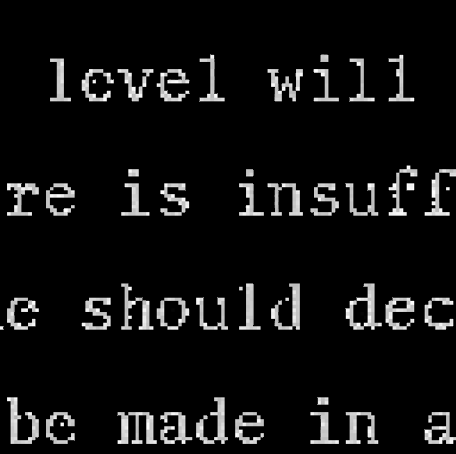}
\caption{Original ``Note" image.}
\label{app:noteOrg}
\end{center}
\end{figure}

\begin{figure}[ht!]
\begin{center}
\noindent
\includegraphics[width=70mm]{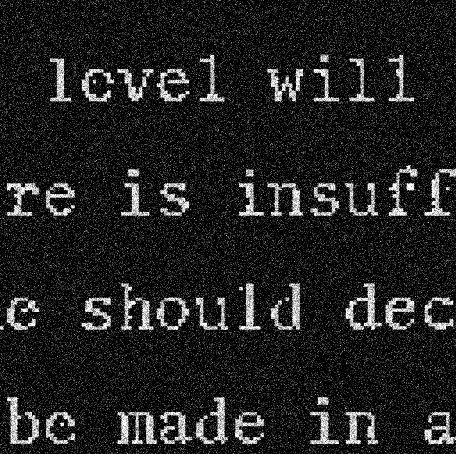}
\caption{``Note" image corrupted with Gaussian noise with $\lambda = 45$.}
\label{app:noteNoise}
\end{center}
\end{figure}

\begin{figure}[ht!]
\begin{center}
\noindent
\includegraphics[width=70mm]{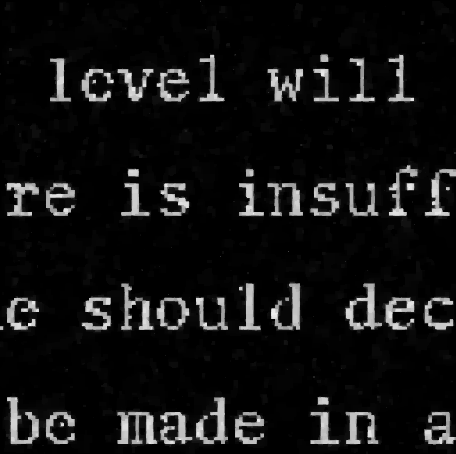}
\caption{Denoised image ``Note" image, using [1 -1] filter; SNR = 15.08.}
\label{app:noteDenoiseOur}
\end{center}
\end{figure}

\begin{figure}[ht!]
\begin{center}
\noindent
\includegraphics[width=70mm]{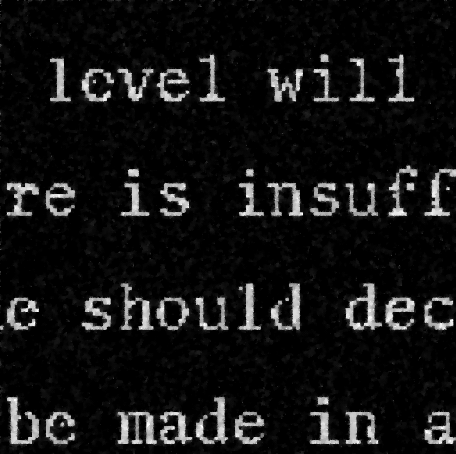}
\caption{Denoised image ``Note" image, using Chambolle's algorithm; SNR = 12.78.}
\label{app:noteDenoiseCh}
\end{center}
\end{figure}

Consider the ``Note" image shown in Fig. \ref{app:noteOrg}. This is corrupted by a zero mean Gaussian noise with $\lambda = 45$ in Fig.~\ref{app:noteNoise}. The image is restored using our method and Chombolle's algorithm \cite{Chambolle} and the denoised images are shown in Fig.~\ref{app:noteDenoiseOur} and \ref{app:noteDenoiseCh}, respectively. The $\lambda$ parameter in (\ref{app:eq:cost}) is manually adjusted to get the best possible results. Our algorithm not only produce a higher SNR, but also a visually better looking image. Solution results for other SNR levels are presented in Table \ref{tab:3}. We also corrupted this image with $\epsilon$-contaminated Gaussian noise (``salt-and-pepper noise"). Denoising results are summarized in Table \ref{tab:4}.

In Table \ref{tab:other}, denoising results for 10 other images with different noise levels are presented. In almost all cases our method produces higher SNR results than the denoising results obtained using \cite{Chambolle}.

\begin{table}[ht!]
\begin{center}
\caption{Comparison of The Results For Denoising Algorithms With Gaussian Noise For Note Image}
\label{tab:3}
\begin{tabular}{|c|c|c|c|}
\hline
\parbox[t]{0.8cm}{Noise \\ std }&\parbox[t]{1cm}{Input \\ SNR}&\parbox[t]{1cm}{\textbf{POCS}}&\parbox[t]{1.5cm}{\textbf{Chambolle}}\\\hline\hline
5&21.12&30.63&29.48\\\hline
10&15.12&25.93&24.20\\\hline
15&11.56&22.91&21.05\\\hline
20&9.06&20.93&18.90\\\hline
25&7.14&19.27&17.17\\\hline
30&5.59&17.89&15.78\\\hline
35&4.21&16.68&14.69\\\hline
40&3.07&15.90&13.70\\\hline
45&2.05&15.08&12.78\\\hline
50&1.12&14.25&12.25\\\hline
\end{tabular}
\end{center}
\end{table}

\begin{table}[ht!]
\begin{center}
\caption{Comparison of The Results For Denoising Algorithms for $\epsilon$-Contamination Noise For Note Image}
\label{tab:4}
\begin{tabular}{|c|c|c|c|c|c|}
\hline
$\epsilon$&$\sigma_{1}$&$\sigma_{2}$&\parbox[t]{1cm}{Input \\ SNR}&\parbox[t]{1cm}{\textbf{POCS}}&\parbox[t]{1.5cm}{\textbf{Chambolle}}\\\hline\hline
0.9&5&30&14.64&23.44&20.56\\\hline
0.9&5&40&12.55&21.39&17.60\\\hline
0.9&5&50&10.75&19.49&15.54\\\hline
0.9&5&60&9.29&17.61&13.82\\\hline
0.9&5&70&7.98&16.01&12.57\\\hline
0.9&5&80&6.89&14.54&11.37\\\hline\hline
0.9&10&30&12.56&22.88&19.74\\\hline
0.9&10&40&11.13&21.00&15.30\\\hline
0.9&10&50&9.85&19.35&12.47\\\hline
0.9&10&60&8.58&17.87&10.42\\\hline
0.9&10&70&7.52&16.38&8.76\\\hline
0.9&10&80&6.46&15.05&7.45\\\hline\hline
0.95&5&30&16.75&24.52&23.18\\\hline
0.95&5&40&14.98&22.59&20.44\\\hline
0.95&5&50&13.41&20.54&18.45\\\hline
0.95&5&60&12.10&18.72&16.80\\\hline
0.95&5&70&10.80&17.13&15.34\\\hline
0.95&5&80&9.76&15.63&14.11 \\\hline\hline
0.95&10&30&13.68&23.79&20.43\\\hline
0.95&10&40&12.66&22.09&15.35\\\hline
0.95&10&50&11.71&20.65&12.28\\\hline
0.95&10&60&10.72&19.10&10.22\\\hline
0.95&10&70&9.82&17.59&8.66\\\hline
0.95&10&80&8.92&16.12&7.34\\\hline
\end{tabular}
\end{center}
\end{table}

\begin{table}[ht!]
\begin{center}
\caption{Comparison of The Results For Denoising Algorithms With Gaussian Noise For Different Images With std = 30, 50}
\label{tab:other}
\begin{tabular}{|c|c|c|c|c|}
\hline
\parbox[t]{0.8cm}{Images }&\parbox[t]{0.7cm}{Noise \\ std}&\parbox[t]{0.8cm}{Input \\ SNR}&\parbox[t]{1cm}{\textbf{POCS}}&\parbox[t]{1.5cm}{\textbf{Chambolle}}\\\hline\hline
House&30&13.85&27.43&27.13\\\hline
House&50&9.45&24.20&24.36\\\hline
Lena&30&12.95&23.63&23.54\\\hline
Lena&50&8.50&21.46&21.37\\\hline
Mandrill&30&13.04&19.98&19.64\\\hline
Mandrill&50&8.61&17.94&17.92\\\hline
Living room&30&12.65&21.21&20.88\\\hline
Living room&50&8.20&19.25&19.05\\\hline
Lake&30&13.44&22.19&21.86\\\hline
Lake&50&8.97&20.03&19.90\\\hline
Jet plane&30&15.57&26.28&25.91\\\hline
Jet plane&50&11.33&23.91&23.54\\\hline
Peppers&30&12.65&23.57&23.59\\\hline
Peppers&50&8.20&21.48&21.36\\\hline
Pirate &30&12.13&21.39&21.30\\\hline
Pirate &50&7.71&19.37&19.43\\\hline
Cameraman&30&12.97&24.13&23.67\\\hline
Cameraman&50&8.55&21.55&21.22\\\hline
Flower&30&11.84&21.97&20.89\\\hline
Flower&50&7.42&19.00&18.88\\\hline\hline
Average&30&13.11&23.18&22.84\\\hline
Average&50&8.69&20.82&20.70\\\hline
\end{tabular}
\end{center}
\end{table}

\section{Conclusion}
\label{sec:Conclusion}
A new denoising method based on the epigraph of the TV function is developed. The solution is obtained using POCS. The new algorithm does not need the optimization of the regularization parameter.

\begin{figure}[ht]
\centering
\subfigure[House]
{\label{fig:house}\includegraphics[width=40mm]{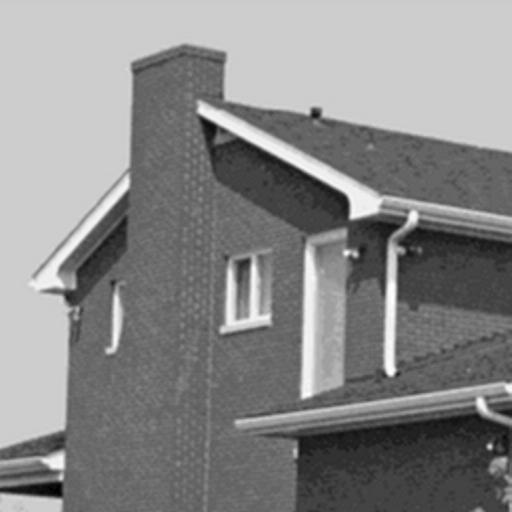}} \quad
\subfigure[Jet plane]
{\label{fig:jetplane}\includegraphics[width=40mm]{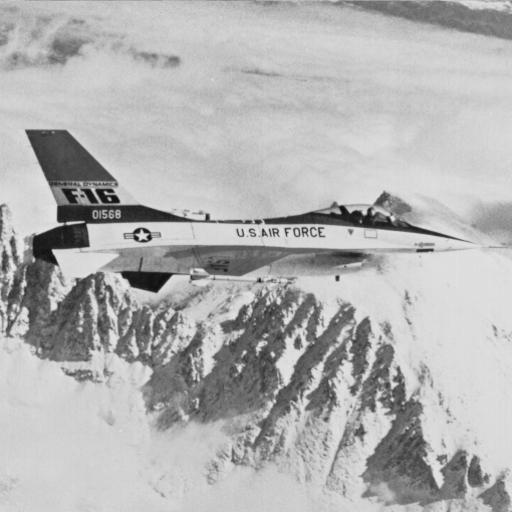}} \qquad
\subfigure[Lake]
{\label{fig:lake}\includegraphics[width=40mm]{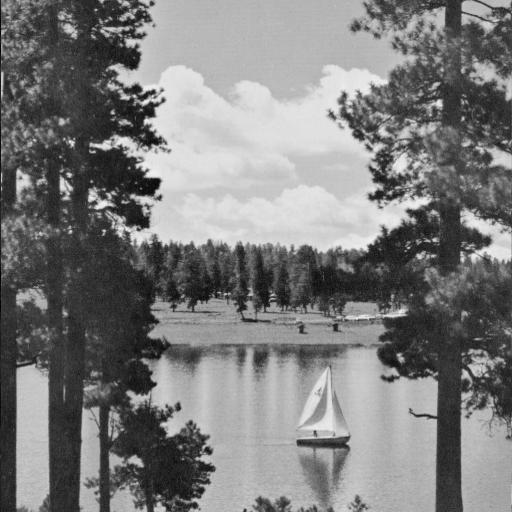}} \quad
\subfigure[Lena]
{\label{fig:lena}\includegraphics[width=40mm]{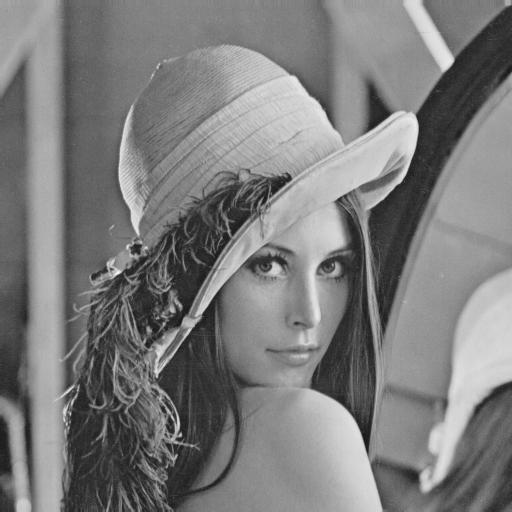}} \qquad
\subfigure[Living room]
{\label{fig:livingroom}\includegraphics[width=40mm]{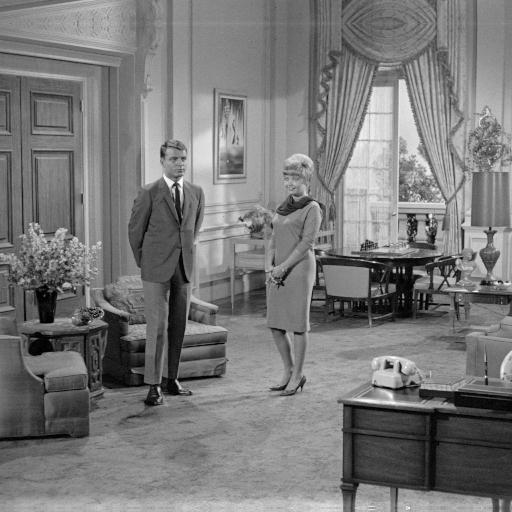}} \quad
\subfigure[Mandrill]
{\label{fig:mandril}\includegraphics[width=40mm]{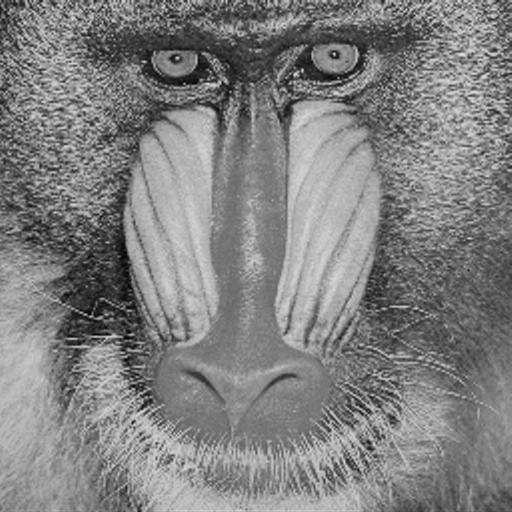}}
\subfigure[Peppers]
{\label{fig:peppers}\includegraphics[width=40mm]{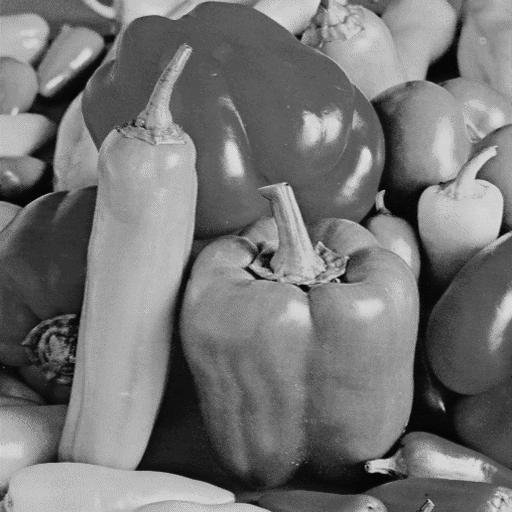}} \quad
\subfigure[Pirate]
{\label{fig:pirate}\includegraphics[width=40mm]{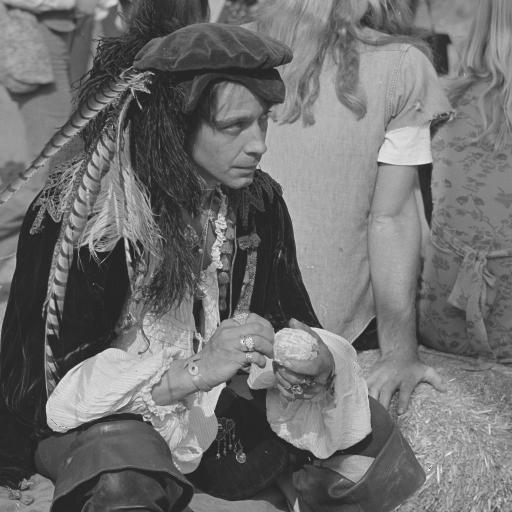}}
\subfigure[Flower]
{\label{fig:peppers}\includegraphics[width=40mm]{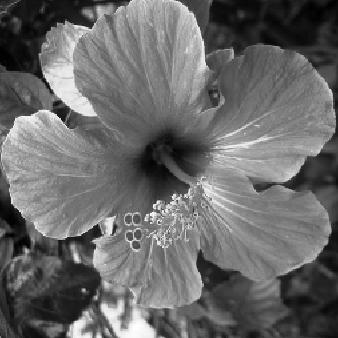}} \quad
\subfigure[Cameraman]
{\label{fig:pirate}\includegraphics[width=40mm]{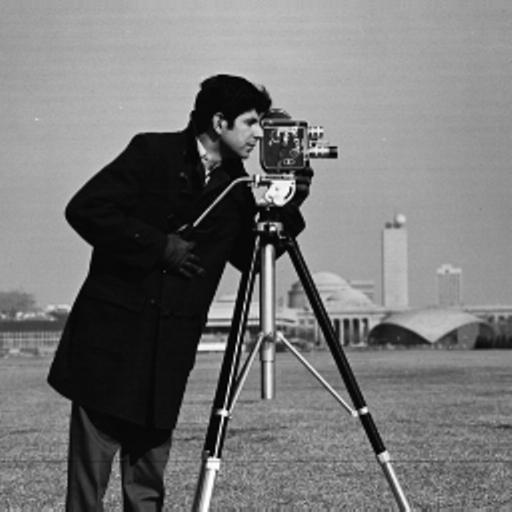}}
\caption{Sample images used in our experiments (a) House, (b) Jet plane, (c) Lake, (d) Lena, (e) Living room, (f) Mandrill, (g) Peppers, (h) Pirate.}
\label{fig:otherOrg}
\end{figure}

\clearpage
\bibliographystyle{IEEEtran}
\bibliography{PhdReferences}






@article{Ach07,
 author               = {Achtman, Neil and Afsheen Afshar and Gopal Santhanam and Byron M Yu and Stephen I Ryu and Krishna V Shenoy},
 journal              = {Journal of Neural Engineering},
 number               = {3},
 pages                = {336},
 title                = {Free-{P}aced {H}igh-{P}erformance {B}rain-{C}omputer {I}nterfaces},
 volume               = {4},
 year                 = {2007},
 }

@article{Ach08,
 author               = {Acharya, S. and F. Tenore and V. Aggarwal and R. Etienne-Cummings and M.H. Schieber and N.V. Thakor},
 doi                  = {10.1109/TNSRE.2007.916269},
 issn                 = {1534-4320},
 journal              = {Neural Systems and Rehabilitation Engineering, IEEE Transactions on},
 keywords             = {ANN;artificial neural network based filters;brain-machine interface;decoding accuracy;dexterous control;dexterous finger movements;individuated finger movements decoding;microelectrode array architectures;primary motor M1 hand area;random neuron subpopulations;rhesus monkeys;single-unit activities recording;volume-constrained neuronal ensembly;wrist movements;arrays;biocontrol;bioelectric phenomena;biomechanics;biomedical measurement;encoding;filtering theory;man-machine systems;mechanoception;medical computing;microelectrodes;neural nets;neurophysiology;user interfaces;Algorithms;Animals;Computer Simulation;Conditioning, Operant;Efferent Pathways;Fingers;Hand;Macaca mulatta;Male;Microelectrodes;Models, Statistical;Motor Cortex;Motor Neurons;Movement;Nonlinear Dynamics;Prosthesis Design;Wrist;},
 month                = {February},
 number               = {1},
 pages                = {15 -23},
 title                = {Decoding {I}ndividuated {F}inger {M}ovements {U}sing {V}olume-{C}onstrained {N}euronal {E}nsembles in the {M1} {H}and {A}rea},
 volume               = {16},
 year                 = {2008},
 }

@article{Agg08,
 author               = {Aggarwal, V. and S. Acharya and F. Tenore and Hyun-Chool Shin and R. Etienne-Cummings and M.H. Schieber and N.V. Thakor},
 doi                  = {10.1109/TNSRE.2007.916289},
 issn                 = {1534-4320},
 journal              = {Neural Systems and Rehabilitation Engineering, IEEE Transactions on},
 keywords             = {M1 neurons;asynchronous decoding;brain-machine interfaces;dexterous finger movements;multifingered hand prosthesis;neural decoding;neural interface;neuroprosthesis;rhesus monkeys;biocontrol;biology computing;decoding;medical control systems;neurophysiology;prosthetics;user interfaces;Algorithms;Animals;Artificial Limbs;Electrophysiology;Fingers;Hand;Macaca mulatta;Male;Models, Statistical;Motor Neurons;Motor Skills;Movement;Neural Networks (Computer);Prosthesis Design;Robotics;Wrist;},
 month                = {February},
 number               = {1},
 pages                = {3 -14},
 title                = {Asynchronous {D}ecoding of {D}exterous {F}inger {M}ovements {U}sing {M1} {N}eurons},
 volume               = {16},
 year                 = {2008},
 }

@article{Akg00,
 author               = {Akg{\"u}l, T. and Sun Mingui and R.J. Sclahassi and A.E \c{C}etin},
 issue                = {8},
 journal              = {IEEE Trans. on Biomedical Engineering},
 pages                = {997-1009},
 title                = {Characterization of {S}leep {S}pindles {U}sing {H}igher {O}rder {S}tatistics and {S}pectra},
 volume               = {47},
 year                 = {2000},
 }

@inproceedings{And04,
 author               = {Andersen, R.A. and J.W. Burdick and S. Musallam and H. Scherberger and B. Pesaran and D. Meeker and B.D. Corneil and I. Fineman and Z. Nenadic and E. Branchaud and J.G. Cham and B. Greger and Y.C. Tai and M.M. Mojarradi},
 booktitle            = {Engineering in Medicine and Biology Society, 2004. IEMBS '04. 26th Annual International Conference of the IEEE},
 doi                  = {10.1109/IEMBS.2004.1404494},
 keywords             = {Automatic testing;Automation;Decoding;Electrodes;Extracellular;Neural prosthesis;Neurons;Probes;Prosthetics;System testing;bioelectric potentials;biological tissues;biomedical electrodes;brain;neurophysiology;prosthetics;automation algorithms;cortex;electrodes;local field potentials;monkeys;neural prosthetics;neurons;spike recordings;tissue;LFP;Movable Electrodes;Neural Prosthetics;},
 month                = {sept.},
 pages                = {5352 -5355},
 title                = {Recording {A}dvances for {N}eural {P}rosthetics},
 volume               = {2},
 year                 = {2004},
 }

@article{And04a,
 accessed             = {03-June-2013},
 author               = {Andersen, Richard A and Sam Musallam and Bijan Pesaran},
 doi                  = {10.1016/j.conb.2004.10.005},
 issn                 = {0959-4388},
 journal              = {Current Opinion in Neurobiology},
 number               = {6},
 pages                = {720 - 726},
 title                = {Selecting the {S}ignals for a {B}rainâ€“{M}achine {I}nterface},
 url                  = {},
 volume               = {14},
 year                 = {2004},
 }

@article{And98,
 author               = {Anderson, C.W. and E.A. Stolz and S. Shamsunder},
 doi                  = {10.1109/10.661153},
 issn                 = {0018-9294},
 journal              = {Biomedical Engineering, IEEE Transactions on},
 keywords             = {0.25 s;EEG analysis;Karhunen-Loeve transform;correlation matrix;device control;eigenvalues;error backpropagation algorithm;feature vectors;mental tasks;multivariate autoregressive models;paralyzed persons;scalar model coefficients;six-channel EEG;spontaneous electroencephalographic signals classification;wheelchair;electroencephalography;feedforward neural nets;medical signal processing;physiological models;Electroencephalography;Feasibility Studies;Humans;Mental Processes;Models, Statistical;Multivariate Analysis;Neural Networks (Computer);Regression Analysis;},
 month                = {March},
 number               = {3},
 pages                = {277 -286},
 title                = {Multivariate {A}utoregressive {M}odels for {C}lassification of {S}pontaneous {E}lectroencephalographic {S}ignals {D}uring {M}ental {T}asks},
 volume               = {45},
 year                 = {1998},
 }

@article{Arv11,
 author               = {Arvaneh, M. and Cuntai Guan and Kai Keng Ang and Chai Quek},
 doi                  = {10.1109/TBME.2011.2131142},
 issn                 = {0018-9294},
 journal              = {Biomedical Engineering, IEEE Transactions on},
 keywords             = {Fisher criterion;brain-computer interfaces;motor imagery datasets;multichannel EEG;optimization;severe motor disabilities;signal classification;sparse common spatial pattern algorithm;support vector machine;brain-computer interfaces;data analysis;electroencephalography;medical disorders;medical signal processing;neurophysiology;signal classification;support vector machines;},
 month                = {June},
 number               = {6},
 pages                = {1865 -1873},
 title                = {Optimizing the {C}hannel {S}election and {C}lassification {A}ccuracy in {EEG}-{B}ased {BCI}},
 volume               = {58},
 year                 = {2011},
 }

@article{Bal09,
 author               = {Ball, T. and M. Kern and I. Mutschler and A. Aertsen and A. Schulze-Bonhage},
 journal              = {Neuroimage},
 number               = {3},
 pages                = {708--716},
 publisher            = {Elsevier},
 title                = {Signal {Q}uality of {S}imultaneously {R}ecorded {I}nvasive and {N}on-{I}nvasive {EEG}},
 volume               = {46},
 year                 = {2009},
 }

@article{Bar07,
 author               = {Baraniuk, R.G.},
 doi                  = {10.1109/MSP.2007.4286571},
 issn                 = {1053-5888},
 journal              = {Signal Processing Magazine, IEEE},
 keywords             = {Nyquist criterion;data compression;optimisation;signal reconstruction;signal representation;Nyquist rate;compressive sensing;nonadaptive linear projections;optimization process;signal capturing;signal reconstruction;signal representation;Bandwidth;Data acquisition;Digital cameras;Image coding;Image sampling;Image storage;Matrices;Signal processing;Transform coding;Vectors},
 number               = {4},
 pages                = {118-121},
 title                = {Compressive Sensing [Lecture Notes]},
 volume               = {24},
 year                 = {2007},
 }

@misc{Bci05,
 accessed             = {03-June-2013},
 author               = {Dornhege, Guido and Benjamin Blankertz and Gabriel Curio and Klaus-Robert M\"uller},
 title                = {{BCI} {C}ompetion {III}, {D}ataset {IVa}},
 url                  = {},
 year                 = {2005},
 }

@article{Bir81,
 accessed             = {03-June-2013},
 author               = {Birbaumer, Niels and Thomas Elbert and Brigitte Rockstroh and Werner Lutzenberger},
 issn                 = {00207594},
 journal              = {International Journal of Psychology},
 number               = {1-4},
 pages                = {389 - 415},
 title                = {Biofeedback of {E}vent-{R}elated {S}low {P}otentials of the {B}rain.},
 url                  = {},
 volume               = {16},
 year                 = {1981},
 }

@article{Bla06,
 author               = {Blankertz, B. and K.-R. M\"uller and D.J. Krusienski and G. Schalk and J.R. Wolpaw and A. Schlogl and G. Pfurtscheller and Jd.R. Mill{\'a}n and M. Schr\"oder and N. Birbaumer},
 doi                  = {10.1109/TNSRE.2006.875642},
 issn                 = {1534-4320},
 journal              = {Neural Systems and Rehabilitation Engineering, {IEEE} Transactions on},
 keywords             = {BCI;adaptive controllers;brain activity;brain-computer interface;device control commands;machine learning;pattern classification;signal analysis;electroencephalography;handicapped aids;learning (artificial intelligence);medical control systems;medical signal processing;pattern classification;Algorithms;Brain;Communication Aids for Disabled;Databases, Factual;Electroencephalography;Evoked Potentials;Humans;Neuromuscular Diseases;Software Validation;Technology Assessment, Biomedical;User-Computer Interface;},
 month                = {June},
 number               = {2},
 pages                = {153 -159},
 title                = {The {BCI} {C}ompetition {III}: {V}alidating {A}lternative {A}pproaches to {A}ctual {BCI} {P}roblems},
 volume               = {14},
 year                 = {2006},
 }

@article{Bla08,
 author               = {Blankertz, B. and R. Tomioka and S. Lemm and M. Kawanabe and K.-R. M\"uller},
 journal              = {Signal Processing Magazine, IEEE},
 number               = {1},
 pages                = {41 -56},
 title                = {Optimizing {S}patial {F}ilters for {R}obust {EEG} {S}ingle-{T}rial {A}nalysis},
 volume               = {25},
 year                 = {2008},
 }

@book{Boy04,
 accessed             = {03-June-2013},
 author               = {Boyd, Stephen and Lieven Vandenberghe},
 howpublished         = {Hardcover},
 isbn                 = {0521833787},
 keywords             = {machine learning},
 month                = {March},
 publisher            = {Cambridge University Press},
 title                = {Convex {O}ptimization},
 url                  = {},
 year                 = {2004},
 }

@article{Bre67,
 accessed             = {03-June-2013},
 author               = {Bregman, L.M.},
 doi                  = {10.1016/0041-5553(67)90040-7},
 issn                 = {0041-5553},
 journal              = {\{USSR\} Computational Mathematics and Mathematical Physics},
 number               = {3},
 pages                = {200 - 217},
 title                = {The {R}elaxation {M}ethod of {F}inding the {C}ommon {P}oint of {C}onvex {S}ets and {I}ts {A}pplication to the {S}olution of {P}roblems in {C}onvex {P}rogramming},
 url                  = {},
 volume               = {7},
 year                 = {1967},
 }

@article{Bregman,
 accessed             = {03-June-2013},
 author               = {Bregman, L.M.},
 doi                  = {10.1016/0041-5553(67)90040-7},
 issn                 = {0041-5553},
 journal              = {\{USSR\} Dokl. Akad. Nauk SSSR},
 number               = {3},
 pages                = {200 - 217},
 title                = {Finding the common point of convex sets by the method of successive projection.(Russian)},
 url                  = {},
 volume               = {7},
 year                 = {1965},
 }

@book{Bur05,
 author               = {Burden, R.L. and J.D. Faires},
 chapter              = {2.1},
 edition              = {8},
 pages                = {48-50},
 publisher            = {Thomson Brooks/ Cole},
 title                = {Numerical {A}nalysis},
 year                 = {2005},
 }

@article{Cad02,
 accessed             = {03-June-2013},
 author               = {Cadzow, James A.},
 doi                  = {10.1006/dspr.2001.0409},
 issn                 = {1051-2004},
 journal              = {Digital Signal Processing},
 keywords             = {inconsistent linear equations.},
 number               = {4},
 pages                = {524 - 560},
 title                = {Minimum $\ell_1$, $\ell_2$, and $\ell_\infty$ {N}orm {A}pproximate {S}olutions to an {O}verdetermined {S}ystem of {L}inear {E}quations},
 url                  = {},
 volume               = {12},
 year                 = {2002},
 }

@article{Can08,
 author               = {Candes, E.J. and Wakin, M.B.},
 doi                  = {10.1109/MSP.2007.914731},
 issn                 = {1053-5888},
 journal              = {Signal Processing Magazine, IEEE},
 keywords             = {data acquisition;image processing;signal processing equipment;signal sampling;Relatively few wavelet;compressed sensing;compressive sampling;data acquisition;image recovery;sampling paradigm;sensing paradigm;signal recovery;Biomedical imaging;Data acquisition;Frequency;Image coding;Image sampling;Protocols;Receivers;Sampling methods;Signal processing;Signal sampling},
 number               = {2},
 pages                = {21-30},
 title                = {An Introduction To Compressive Sampling},
 volume               = {25},
 year                 = {2008},
 }

@article{Cen12,
 accessed             = {03-June-2013},
 author               = {Censor, Yair and Wei Chen and Patrick L. Combettes and Ran Davidi and GaborT. Herman},
 doi                  = {10.1007/s10589-011-9401-7},
 issn                 = {0926-6003},
 journal              = {Computational Optimization and Applications},
 keywords             = {Projection methods; Convex feasibility problems; Numerical evaluation; Optimization; Linear inequalities; Sparse matrices},
 language             = {English},
 number               = {3},
 pages                = {1065-1088},
 publisher            = {Springer US},
 title                = {On the {E}ffectiveness of {P}rojection {M}ethods for {C}onvex {F}easibility {P}roblems with {L}inear {I}nequality {C}onstraints},
 url                  = {},
 volume               = {51},
 year                 = {2012},
 }

@article{Cen81,
 author               = {Censor, Yair and Arnold Lent},
 journal              = {Journal of Optimization Theory and Applications},
 number               = {3},
 pages                = {321-353},
 title                = {An {I}terative {R}ow-{A}ction {M}ethod for {I}nterval {C}onvex {P}rogramming},
 volume               = {34},
 year                 = {1981},
 }

@article{Cen82,
 accessed             = {05-June-2013},
 author               = {Censor, Yair and Tommy Elfving},
 doi                  = {10.1016/0024-3795(82)90149-5},
 issn                 = {0024-3795},
 journal              = {Linear Algebra and its Applications},
 number               = {0},
 pages                = {199 - 211},
 title                = {New {M}ethods for {L}inear {I}nequalities},
 url                  = {},
 volume               = {42},
 year                 = {1982},
 }

@article{Cet03,
 author               = {\c{C}etin, A.E. and H. \"Ozakta\c{s} and H.M. Ozaktas},
 doi                  = {10.1049/el:20031119},
 issn                 = {0013-5194},
 journal              = {Electronics Letters},
 keywords             = {Fourier transforms;interpolation;iterative methods;signal reconstruction;signal representation;signal resolution;POCS algorithm;discrete spaces;iterative algorithm;low resolution sensors;low resolution wavefields;partial fractional Fourier transform information;projections onto convex sets;resolution enhancement;signal interpolation;space-domain signal samples},
 number               = {25},
 pages                = {1808-1810},
 title                = {Resolution {E}nhancement of {L}ow {R}esolution {W}avefields with},
 volume               = {39},
 year                 = {2003},
 }

@article{Cet97,
 author               = {A. E. \c{C}etin and O.N. Gerek and Y. Yardimci},
 doi                  = {10.1109/79.581378},
 issn                 = {1053-5888},
 journal              = {IEEE Signal Processing Magazine},
 keywords             = {FIR filters;delay circuits;digital filters;discrete time filters;fast Fourier transforms;filtering theory;frequency response;iterative methods;multidimensional digital filters;FFT algorithm;FIR discrete-time filters;Parks-McClellan algorithm;convexity property;equiripple FIR filter design;equiripple approximation;fast Fourier transform;frequency domain constraints;frequency response;image processing;impulse response support;iterative design method;linear phase FIR filter;linear programming;magnitude response bounds;multidimensional FFT computations;multidimensional FIR filter;signal processing;time domain constraints;zero-phase FIR filter;Algorithm design and analysis;Design methodology;Fast Fourier transforms;Finite impulse response filter;Frequency domain analysis;Frequency response;Image processing;Iterative algorithms;Linear programming;Signal processing},
 number               = {2},
 pages                = {60-64},
 title                = {Equiripple {FIR} {F}ilter {D}esign by the {FFT} {A}lgorithm},
 volume               = {14},
 year                 = {1997},
 }

@misc{Cha01,
 accessed             = {03-June-2013},
 author               = {Chang, Chih-Chung and Chih-Jen Lin},
 title                = {A {L}ibrary for {S}upport {V}ector {M}achines},
 url                  = {},
 year                 = {2001},
 }

@article{Cha10,
 author               = {Chao, Z.C. and Y. Nagasaka and N. Fujii},
 journal              = {Frontiers in neuroengineering},
 publisher            = {Frontiers Research Foundation},
 title                = {Long-{T}erm {A}synchronous {D}ecoding of {A}rm {M}otion {U}sing {E}lectrocorticographic {S}ignals in {M}onkeys},
 volume               = {3},
 year                 = {2010},
 }

@article{Cha99,
 author               = {Chapin, J.K. and K.A. Moxon and R.S. Markowitz and M.A.L. Nicolelis},
 journal              = {Nature neuroscience},
 pages                = {664--670},
 publisher            = {NATURE AMERICA},
 title                = {Real-{T}ime {C}ontrol of a {R}obot {A}rm {U}sing {S}imultaneously {R}ecorded {N}eurons in the {M}otor {C}ortex},
 volume               = {2},
 year                 = {1999},
 }

@article{Com04,
 author               = {Patrick L. Combettes and {Jean-Christophe} Pesquet},
 journal              = {IEEE Transactions on Image Processing},
 pages                = {1213--1222},
 title                = {Image restoration subject to a total variation constraint},
 volume               = {13},
 year                 = {2004},
 }

@incollection{Com11,
 author               = {Combettes, Patrick L. and Pesquet, Jean-Christophe},
 booktitle            = {Fixed-Point Algorithms for Inverse Problems in Science and Engineering},
 doi                  = {10.1007/978-1-4419-9569-8_10},
 editor               = {Bauschke, Heinz H. and Burachik, Regina S. and Combettes, Patrick L. and Elser, Veit and Luke, D. Russell and Wolkowicz, Henry},
 isbn                 = {978-1-4419-9568-1},
 keywords             = {Alternating-direction method of multipliers; Backwardâ€“backward algorithm; Convex optimization; Denoising; Douglasâ€“Rachford algorithm; Forwardâ€“backward algorithm; Frame; Landweber method; Iterative thresholding; Parallel computing; Peacemanâ€“Rachford algorithm; Proximal algorithm; Restoration and reconstruction; Sparsity; Splitting},
 language             = {English},
 pages                = {185-212},
 publisher            = {Springer New York},
 series               = {Springer Optimization and Its Applications},
 title                = {Proximal Splitting Methods in Signal Processing},
 url                  = {},
 year                 = {2011},
 }

@misc{Com12,
 author               = {Combettes, Patrick L.},
 title                = {Algorithmes proximaux pour l\'{e}s problemes d\'{ }optimisation structur les},
 url                  = {},
 year                 = {2012},
 }

@article{Com93,
 author               = {Combettes, P.L.},
 doi                  = {10.1109/5.214546},
 issn                 = {0018-9219},
 journal              = {Proceedings of the IEEE},
 keywords             = {analytical techniques;feasibility;set theoretic estimation;signal processing;solution space;systems science;estimation theory;set theory;},
 month                = {February},
 number               = {2},
 pages                = {182 -208},
 title                = {The foundations of set theoretic estimation},
 volume               = {81},
 year                 = {1993},
 }

@article{Das97,
 accessed             = {03-June-2013},
 author               = {Dash, M. and H. Liu},
 journal              = {Intelligent Data Analysis},
 pages                = {131--156},
 title                = {Feature {S}election for {C}lassification},
 url                  = {},
 volume               = {1},
 year                 = {1997},
 }

@book{Dem97,
 author               = {Demmel, J.},
 pages                = {178-179},
 publisher            = {SIAM},
 title                = {Applied {N}umerical {L}inear {A}lgebra},
 year                 = {1997},
 }

@article{Die95,
 author               = {Dietterich, Thomas G. and Ghulum Bakiri},
 journal              = {Journal of Artificial Intelligence Research},
 pages                = {263--286},
 title                = {Solving {M}ulticlass {L}earning {P}roblems Via {E}rror-{C}orrecting {O}utput {C}odes},
 volume               = {2},
 year                 = {1995},
 }

@inproceedings{Dor05,
 author               = {Dornhege, Guido and Benjamin Blankertz and Matthias Krauledat and Florian Losch and Gabriel Curio and Klaus-Robert M\"uller},
 booktitle            = {in Advances in Neural Inf. Proc. Systems (NIPS 05},
 pages                = {315--322},
 publisher            = {MIT Press},
 title                = {Optimizing {S}patio-{T}emporal {F}ilters for {I}mproving {B}rain-{C}omputer {I}nterfacing},
 year                 = {2005},
 }

@article{Dor06,
 author               = {Dornhege, G. and B. Blankertz and M. Krauledat and F. Losch and G. Curio and K.-R. M\"uller},
 doi                  = {10.1109/TBME.2006.883649},
 issn                 = {0018-9294},
 journal              = {Biomedical Engineering, IEEE Transactions on},
 keywords             = {brain-computer interfacing;multichannel EEG single-trials;optimization;paralyzed patients;single-trial brain signal classification;source localization;spatial filters;temporal filters;electroencephalography;handicapped aids;medical signal processing;optimisation;signal classification;spatial filters;},
 month                = {November},
 number               = {11},
 pages                = {2274 -2281},
 title                = {Combined {O}ptimization of {S}patial and {T}emporal {F}ilters for {I}mproving {B}rain-{C}omputer {I}nterfacing},
 volume               = {53},
 year                 = {2006},
 }

@inproceedings{Duc08,
 accessed             = {03-June-2013},
 address              = {New York, NY, USA},
 author               = {Duchi, John and Shai Shalev-Shwartz and Yoram Singer and Tushar Chandra},
 booktitle            = {Proceedings of the 25th international conference on Machine learning},
 doi                  = {10.1145/1390156.1390191},
 isbn                 = {978-1-60558-205-4},
 location             = {Helsinki, Finland},
 numpages             = {8},
 pages                = {272--279},
 publisher            = {ACM},
 series               = {ICML '08},
 title                = {Efficient {P}rojections {O}nto the $\ell_1$-{B}all for {L}earning in {H}igh {D}imensions},
 url                  = {},
 year                 = {2008},
 }

@article{Duc88,
 author               = {Duchene, J. and S. Leclercq},
 doi                  = {10.1109/34.9121},
 issn                 = {0162-8828},
 journal              = {Pattern Analysis and Machine Intelligence, IEEE Transactions on},
 keywords             = {discriminant analysis;discriminant vectors;multivariate data sets;optimal transformation;principal component analysis;computerised pattern recognition;vectors;},
 month                = {November},
 number               = {6},
 pages                = {978 -983},
 title                = {An {O}ptimal {T}ransformation for {D}iscriminant and {P}rincipal {C}omponent {A}nalysis},
 volume               = {10},
 year                 = {1988},
 }

@book{Dud00,
 author               = {Duda, Richard O. and Peter E. Hart and David G. Stork},
 edition              = {Second},
 isbn                 = {0471056693},
 publisher            = {Wiley-Interscience},
 title                = {Pattern {C}lassification},
 year                 = {2000},
 }

@misc{Eeg20,
 accessed             = {03-June-2013},
 title                = {{EEG} 10-20 {S}ystem {D}iagram.},
 url                  = {http://www.medicine.mcgill.ca/physio/vlab/biomed_signals/images/10-20_sys.gif},
 year                 = {2012},
 }

@inproceedings{Far06,
 author               = {Farquhar, J. and N. J. Hill and T. N. Lal and B. Schlkopf},
 booktitle            = {In Proceedings of the 3rd International Brain-Computer Interface Workshop and Training Course},
 title                = {Regularised {CSP} for {S}ensor {S}election in {BCI}},
 year                 = {2006},
 }

@article{Far88,
 accessed             = {03-June-2013},
 author               = {Farwell, L.A. and E. Donchin},
 doi                  = {10.1016/0013-4694(88)90149-6},
 issn                 = {0013-4694},
 journal              = {Electroencephalography and Clinical Neurophysiology},
 keywords             = {EEG},
 number               = {6},
 pages                = {510 - 523},
 title                = {Talking {O}ff the {T}op of {Y}our {H}ead: Toward a {M}ental {P}rosthesis {U}tilizing {E}vent-{R}elated {B}rain {P}otentials},
 url                  = {},
 volume               = {70},
 year                 = {1988},
 }

@article{Fla11,
 author               = {Flamary, R{\'e}mi and Alain Rakotomamonjy},
 journal              = {CoRR},
 title                = {Decoding {F}inger {M}ovements From {ECoG} {S}ignals {U}sing {S}witching {L}inear {M}odels},
 volume               = {abs/1106.3395},
 year                 = {2011},
 }

@inproceedings{Fli12,
 author               = {Flint, R.D. and Z.A. Wright and M.W. Slutzky},
 booktitle            = {Engineering in Medicine and Biology Society (EMBC), 2012 Annual International Conference of the IEEE},
 doi                  = {10.1109/EMBC.2012.6347536},
 issn                 = {1557-170X},
 keywords             = {Correlation;Decoding;Educational institutions;Kinematics;Measurement;Neuroscience;USA Councils;bioelectric potentials;biomimetics;brain-computer interfaces;decoding;medical computing;medical control systems;neurophysiology;2D cursor control;BMI decoder;LFP signals;biomimetic BMI;biomimetic brain machine interface control;cursor movement;cursor position control;local field potentials;random-pursuit reaching task;severe motor impairments;single-unit based BMI;single-unit spikes;static decoder;target acquisition rates;},
 month                = {28 2012-sept. 1},
 pages                = {6719 -6722},
 title                = {Control of a {B}iomimetic {B}rain {M}achine {I}nterface with {L}ocal {F}ield {P}otentials: {P}erformance and {S}tability of a {S}tatic {D}ecoder over 200 {D}ays},
 year                 = {2012},
 }

@article{Ger99,
 accessed             = {03-June-2013},
 author               = {M\"{u}ller-Gerking, Johannes and Gert Pfurtscheller and Henrik Flyvbjerg},
 doi                  = {10.1016/S1388-2457(98)00038-8},
 issn                 = {1388-2457},
 journal              = {Clinical Neurophysiology},
 keywords             = {Event-related desynchronization},
 number               = {5},
 pages                = {787 - 798},
 title                = {Designing {O}ptimal {S}patial {F}ilters for {S}ingle-{T}rial {EEG} {C}lassification in a {M}ovement {T}ask},
 url                  = {},
 volume               = {110},
 year                 = {1999},
 }

@inproceedings{Gok11,
 author               = {G\"{o}ksu, Fikri and Nuri F{\i}rat {\.I}nce and A.H. Tewfik},
 booktitle            = {Acoustics, Speech and Signal Processing (ICASSP), 2011 IEEE International Conference on},
 month                = {May},
 pages                = {533 -536},
 title                = {Sparse {C}ommon {S}patial {P}atterns in {B}rain {C}omputer {I}nterface {A}pplications},
 year                 = {2011},
 }

@inproceedings{Gok11a,
 author               = {G\"{o}ksu, Fikri and F{\i}rat {\.I}nce and {\.I}brahim Onaran},
 booktitle            = {Signals, Systems and Computers (ASILOMAR), 2011 Conference Record of the Forty Fifth Asilomar Conference on},
 month                = {November},
 pages                = {117 -121},
 title                = {Sparse {C}ommon {S}patial {P}atterns with {R}ecursive {W}eight {E}limination},
 year                 = {2011},
 }

@article{Gub67,
 accessed             = {05-June-2013},
 author               = {Gubin, L.G. and B.T. Polyak and E.V. Raik},
 doi                  = {10.1016/0041-5553(67)90113-9},
 issn                 = {0041-5553},
 journal              = {\{USSR\} Computational Mathematics and Mathematical Physics},
 number               = {6},
 pages                = {1 - 24},
 title                = {The {M}ethod of {P}rojections for {F}inding the {C}ommon {P}oint of {C}onvex {S}ets},
 url                  = {},
 volume               = {7},
 year                 = {1967},
 }

@article{Gug00,
 author               = {Guger, C. and H. Ramoser and G. Pfurtscheller},
 doi                  = {10.1109/86.895947},
 issn                 = {1063-6528},
 journal              = {Rehabilitation Engineering, IEEE Transactions on},
 keywords             = {3 d;EEG-based brain-computer interface;amyotrophic lateral sclerosis;brain-computer interface;common spatial patterns;device control;feedback;left-hand motor imagery;on-line discrimination;on-line sessions;real-time EEG analysis;right-hand motor imagery;simple binary response;subject-specific spatial patterns;electroencephalography;handicapped aids;medical signal processing;},
 month                = {December},
 number               = {4},
 pages                = {447 -456},
 title                = {Real-{T}ime {EEG} {A}nalysis with {S}ubject-{S}pecific {S}patial {P}atterns for a {B}rain-{C}omputer {I}nterface ({BCI})},
 volume               = {8},
 year                 = {2000},
 }

@article{Gunay,
 author               = {Osman G\"{u}nay and K{\i}van\c{c} K\"{o}se and Beh\c{et} U\u{g}ur T\"{o}reyin and Ahmet Enis \c{C}etin},
 journal              = {IEEE Transactions on Image Processing},
 keywords             = {Computer science;Convergence of numerical methods;Degradation;Hilbert space;Image reconstruction;Image restoration;Iterative algorithms;Microwave imaging;Sufficient conditions;Vectors},
 number               = {5},
 pages                = {2853-2865},
 title                = {Entropy-functional-based online adaptive decision fusion framework with application to wildfire detection in video},
 volume               = {21},
 year                 = {May 2012},
 }

@article{Guy02,
 author               = {Guyon, Isabelle and Jason Weston and Stephen Barnhill and Vladimir Vapnik},
 journal              = {Machine Learning},
 keywords             = {data, learning, machine, mining, svm},
 number               = {1-3},
 pages                = {389--422},
 publisher            = {Kluwer Academic Publishers, Boston},
 title                = {Gene {S}election for {C}ancer {C}lassification {U}sing {S}upport {V}ector {M}achines},
 volume               = {46},
 year                 = {2002},
 }

@article{Hak09,
 author               = {Tuna, H. and \.{I}. Onaran and A.E. \c{C}etin},
 doi                  = {10.1109/LSP.2009.2024589},
 issn                 = {1070-9908},
 journal              = {Signal Processing Letters, IEEE},
 keywords             = {covariance matrices;image classification;image recognition;image texture;co-difference matrix;covariance method;image classification;image description;multiplier-less operator;texture recognition;Co-difference matrix;covariance matrix;license plate identification;multiplier-less signal processing;texture recognition},
 number               = {9},
 pages                = {751-753},
 title                = {Image {D}escription {U}sing a {M}ultiplier-{L}ess {O}perator},
 volume               = {16},
 year                 = {2009},
 }

@article{Her95,
 author               = {Herman, G. T.},
 journal              = {Real-Time Imaging},
 number               = {1},
 pages                = {3-18},
 title                = {Image {R}econstruction From {P}rojections},
 volume               = {1},
 year                 = {1995},
 }

@incollection{Hil06,
 affiliation          = {MPI for Biological Cybernetics, Spemannstr. 38, 72076 Tbingen},
 author               = {Hill, N. and Thomas Lal and Michael Schr\"oder and Thilo Hinterberger and Guido Widman and Christian Elger and Bernhard Sch\"olkopf and Niels Birbaumer},
 booktitle            = {Pattern Recognition},
 editor               = {Franke, Katrin and M\"uller, Klaus-Robert and Nickolay, Bertram and Sch\"afer, Ralf},
 isbn                 = {978-3-540-44412-1},
 keyword              = {Computer Science},
 pages                = {404-413},
 publisher            = {Springer Berlin / Heidelberg},
 series               = {Lecture Notes in Computer Science},
 title                = {Classifying {E}vent-{R}elated {D}esynchronization in {EEG}, {ECoG} and {MEG} {S}ignals},
 volume               = {4174},
 year                 = {2006},
 }

@article{Hwa09,
 author               = {Hwang, Eun Jung and Richard A. Andersen},
 journal              = {The Journal of Neuroscience},
 month                = {November},
 number               = {45},
 pages                = {14363-14370},
 title                = {Brain {C}ontrol of {M}ovement {E}xecution {O}nset {U}sing {L}ocal {F}ield {P}otentials in {P}osterior {P}arietal {C}ortex},
 volume               = {29},
 year                 = {2009},
 }

@phdthesis{Inc05,
 accessed             = {03-June-2013},
 author               = {{\.I}nce, Nuri F{\i}rat},
 school               = {Cukurova University},
 title                = {Analysis and {C}lassification of {EEG} with {A}dapted {W}avelets and {L}ocal {D}iscriminant {B}ases},
 url                  = {},
 year                 = {2005},
 }

@incollection{Inc09,
 affiliation          = {University of Minnesota Department of Electrical and Computer Engineering 200 Union St. SE Minneapolis MN 55455 U.S.A.},
 author               = {{\.I}nce, Nuri F{\i}rat and Fikri G\"{o}ksu and Ahmed H. Tewfik},
 booktitle            = {Biomedical Engineering Systems and Technologies},
 editor               = {Fred, Ana and Filipe, Joaquim and Gamboa, Hugo},
 keyword              = {Computer Science},
 pages                = {357-374},
 publisher            = {Springer Berlin Heidelberg},
 series               = {Communications in Computer and Information Science},
 title                = {{ECoG} {B}ased {B}rain {C}omputer {I}nterface with {S}ubset {S}election},
 volume               = {25},
 year                 = {2009},
 }

@article{Inc10,
 author               = {{\.I}nce, Nuri F{\i}rat AND Gupta, Rahul AND Arica, Sami AND Tewfik, Ahmed H. AND Ashe, James AND Pellizzer, Giuseppe},
 doi                  = {10.1371/journal.pone.0014384},
 journal              = {PLoS ONE},
 month                = {December},
 number               = {12},
 pages                = {e14384},
 publisher            = {Public Library of Science},
 title                = {High {A}ccuracy {D}ecoding of {M}ovement {T}arget {D}irection in {N}on-{H}uman {P}rimates {B}ased on {C}ommon {S}patial {P}atterns of {L}ocal {F}ield {P}otentials},
 volume               = {5},
 year                 = {2010},
 }

@inproceedings{Jun07,
 accessed             = {03-June-2013},
 author               = {Jung, Sung-yoon and Sung-kyun Kang and Myoung-Jun Lee and Inhyuk Moon},
 booktitle            = {Control, Automation and Systems, 2007. ICCAS '07. International Conference on},
 doi                  = {10.1109/ICCAS.2007.4406884},
 keywords             = {DC motors;dexterous manipulators;motion control;prosthetics;springs (mechanical);DC motor;distal-middle phalange;finger design;hand motion;hand prosthesis;metacarpal bone;proximal phalange;robotic hand;tendon-driven three fingers;Bones;DC motors;Fingers;Grasping;Prosthetics;Robots;Springs;Tendons;Thumb;Wire;finger design;prosthesis;robotic hand;tendon-driven},
 pages                = {83-86},
 title                = {Design of {R}obotic {H}and with {T}endon-{D}riven {T}hree {F}ingers},
 url                  = {},
 year                 = {2007},
 }

@article{Kap11,
 accessed             = {03-June-2013},
 author               = {Kaplan, Robert M},
 journal              = {Australasian Psychiatry},
 number               = {2},
 pages                = {168--169},
 publisher            = {SAGE Publications},
 title                = {The {M}ind {R}eader: the {F}orgotten {L}ife of {H}ans {B}erger, {D}iscoverer of the {EEG}},
 url                  = {},
 volume               = {19},
 year                 = {2011},
 }

@inproceedings{Kee05,
 accessed             = {03-June-2013},
 address              = {New York, NY, USA},
 author               = {Keerthi, S. Sathiya},
 booktitle            = {Proceedings of the 22nd international conference on Machine learning},
 doi                  = {10.1145/1102351.1102404},
 isbn                 = {1-59593-180-5},
 location             = {Bonn, Germany},
 numpages             = {8},
 pages                = {417--424},
 publisher            = {ACM},
 series               = {ICML '05},
 title                = {Generalized {LARS} as an {E}ffective {F}eature {S}election {T}ool for {T}ext {C}lassification with {SVM}s},
 url                  = {},
 year                 = {2005},
 }

@article{Kem08,
 author               = {Kemere, Caleb and Gopal Santhanam and Byron M. Yu and Afsheen Afshar and Stephen I. Ryu and Teresa H. Meng and and Krishna V. Shenoy},
 file                 = {:../../#References/[Kem08] Detecting Neural-State Transitions Using Hidden Markov Models for Motor Cortical Prostheses.pdf:PDF},
 journal              = {Journal of NeuroPhysiology},
 month                = {June},
 number               = {100},
 pages                = {2442-2452},
 title                = {Detecting {N}eural-{S}tate {T}ransitions {U}sing {H}idden {M}arkov {M}odels for {M}otor {C}ortical {P}rostheses},
 year                 = {2008},
 }

@article{Kim92,
 author               = {A.~E.~Cetin and R.~Ansari},
 journal              = {JOSA-A},
 pages                = {673-676},
 title                = {Convolution based framework for signal recovery and applications},
 year                 = {1988},
 }

@article{Cetin94,
 author               = {A.~E.~Cetin and R.~Ansari},
 journal              = {IEEE Transactions on Signal Processing},
 pages                = {673-676},
 title                = {Signal recovery from wavelet transform maxima},
 year                 = {1994},
 }



@article{Kir11,
 author               = {Kirsch, D.B.},
 journal              = {Chest},
 number               = {4},
 pages                = {939--946},
 publisher            = {American College of Chest Physicians},
 title                = {There and {B}ack {A}gain : A {C}urrent {H}istory of {S}leep {M}edicine},
 volume               = {139},
 year                 = {2011},
 }

@phdthesis{Kiv12,
 accessed             = {03-June-2013},
 author               = {K\"{o}se, K{\i}van\c{c}},
 school               = {Bilkent University},
 title                = {Signal and {I}mage {P}rocessing {A}lgorithms {U}sing {I}nterval {C}onvex {P}rogramming and {S}parsity},
 url                  = {},
 year                 = {2012},
 }

@article{Kol90,
 author               = {Koles, Zoltan J. and Michael S. Lazar and Steven Z. Zhou},
 issn                 = {0896-0267},
 issue                = {4},
 journal              = {Brain Topography},
 keyword              = {Biomedical and Life Sciences},
 pages                = {275-284},
 publisher            = {Springer New York},
 title                = {Spatial {P}atterns {U}nderlying {P}opulation {D}ifferences in the {B}ackground {EEG}},
 volume               = {2},
 year                 = {1990},
 }

@article{Kos12,
 author               = {Kose, K. and Cevher, V. and Cetin, A.E.},
 journal              = {Acoustics, Speech and Signal Processing (ICASSP), 2012 IEEE International Conference on},
 doi                  = {10.1109/ICASSP.2012.6288628},
 issn                 = {1520-6149},
 keywords             = {filtering theory;set theory;signal denoising;transforms;convex sets;filtered variation method;sparse signal denoising;sparse signal processing;transform domains;Discrete Fourier transforms;Image restoration;Noise;Noise reduction;TV;Filtered variation;projection onto convex sets;regularization;total variation},
 pages                = {3329-3332},
 title                = {Filtered Variation method for denoising and sparse signal processing},
 year                 = {2012},
 }

@article{Kot97,
 author               = {Kotchoubey, B. and H. Schleichert and W. Lutzenberger and N. Birbaumer},
 journal              = {Applied psychophysiology and biofeedback},
 number               = {2},
 pages                = {77--93},
 publisher            = {Springer},
 title                = {A {N}ew {M}ethod for {S}elf-{R}egulation of {S}low {C}ortical {P}otentials in a {T}imed {P}aradigm},
 volume               = {22},
 year                 = {1997},
 }

@article{Kub98,
 author               = {K{\"u}bler, A. and B. Kotchoubey and H.P. Salzmann and N. Ghanayim and J. Perelmouter and V. H{\"o}mberg and N. Birbaumer},
 journal              = {Neuroscience letters},
 number               = {3},
 pages                = {171--174},
 publisher            = {Elsevier},
 title                = {Self-{R}egulation of {S}low {C}ortical {P}otentials in {C}ompletely {P}aralyzed {H}uman {P}atients},
 volume               = {252},
 year                 = {1998},
 }

@article{Lal04,
 author               = {Lal, T.N. and M. Schroder and T. Hinterberger and J. Weston and M. Bogdan and N. Birbaumer and B. Scholkopf},
 doi                  = {10.1109/TBME.2004.827827},
 issn                 = {0018-9294},
 journal              = {Biomedical Engineering, IEEE Transactions on},
 keywords             = {BCI;art feature selection algorithms;brain activity classification;brain computer interface;electroencephalogram signals;feature selection;mental task;motor imagery paradigm;recursive feature elimination;standard filter methods;support vector channel selection;support vector machines;zero-norm optimization;electroencephalography;feature extraction;medical signal processing;signal classification;support vector machines;Algorithms;Artificial Intelligence;Cerebral Cortex;Cluster Analysis;Electroencephalography;Evoked Potentials, Motor;Hand;Humans;Male;Pattern Recognition, Automated;Reproducibility of Results;Sensitivity and Specificity;User-Computer Interface;},
 month                = {June},
 number               = {6},
 pages                = {1003 -1010},
 title                = {Support {V}ector {C}hannel {S}election in {BCI}},
 volume               = {51},
 year                 = {2004},
 }

@inproceedings{Lal05,
 author               = {Lal, Thomas Navin and Thilo Hinterberger and Guido Widman and Christian E. Elger and Bernhard Sch\"olkopf and Niels Birbaumer},
 booktitle            = {in Advances in Neural Information Processing Systems 17},
 pages                = {737--744},
 publisher            = {MIT Press},
 title                = {Methods Towards {I}nvasive {H}uman {B}rain {C}omputer {I}nterfaces},
 year                 = {2005},
 }

@article{Kose11,
 author               = {K. Kose and A. E. Cetin},
 journal              = {IEEE Signal Processing Magazine},
 pages                = {117--121},
 publisher            = {IEEE},
 title                = {Low-Pass Filtering of Irregularly Sampled Signals Using a Set Theoretic Framework},
 year                 = {2011},
 }

@article{Cetin89,
 author               = {A. E. Cetin},
 journal              = {Signal Processing},
 pages                = {129--148},
 publisher            = {Elsevier},
 title                = {Reconstruction of signals from Fourier transform samples},
 year                 = {1989},
 }

@inproceedings{Lee06,
 author               = {Lee, Su-in and Honglak Lee and Pieter Abbeel and Andrew Y. Ng},
 booktitle            = {In AAAI},
 title                = {Efficient {L1} {R}egularized {L}ogistic {R}egression},
 year                 = {2006},
 }

@article{Lem05,
 author               = {Lemm, S. and B. Blankertz and G. Curio and K.-R. M\"uller},
 doi                  = {10.1109/TBME.2005.851521},
 issn                 = {0018-9294},
 journal              = {Biomedical Engineering, IEEE Transactions on},
 keywords             = {EEG classification;brain-computer interface;common spatial pattern;electrode;electroencephalogram;frequency filters;imagined limb movements;information transfer rate;machine learning;single trial EEG;spatio-spectral filters;time delay embedding;biomechanics;biomedical electrodes;electroencephalography;handicapped aids;learning (artificial intelligence);medical signal processing;optical filters;signal classification;spatial filters;Brain;Diagnosis, Computer-Assisted;Electroencephalography;Evoked Potentials;Humans;Pattern Recognition, Automated;Reproducibility of Results;Sensitivity and Specificity;User-Computer Interface;},
 month                = {September},
 number               = {9},
 pages                = {1541 -1548},
 title                = {Spatio-{S}pectral {F}ilters for {I}mproving the {C}lassification of {S}ingle {T}rial {EEG}},
 volume               = {52},
 year                 = {2005},
 }

@incollection{Les09,
 author               = {Mackey, Lester},
 booktitle            = {Advances in Neural Information Processing Systems 21},
 editor               = {D. Koller and D. Schuurmans and Y. Bengio and L. Bottou},
 pages                = {1017--1024},
 title                = {Deflation {M}ethods for {S}parse {PCA}},
 year                 = {2009},
 }

@article{Leu04,
 author               = {Leuthardt, E.C. and G. Schalk and J.R. Wolpaw and J.G. Ojemann and D.W. Moran},
 journal              = {Journal of Neural Engineering},
 pages                = {63},
 publisher            = {IOP Publishing},
 title                = {A {B}rain-{C}omputer {I}nterface {U}sing {E}lectrocorticographic {S}ignals in {H}umans},
 volume               = {1},
 year                 = {2004},
 }

@article{Lev99,
 author               = {Levine, S P},
 issue                = {5},
 journal              = {Journal of clinical neurophysiology},
 pages                = {439},
 title                = {Identification of {E}lectrocorticogram {P}atterns as the {B}asis for a {D}irect {B}rain {I}nterface},
 volume               = {16},
 year                 = {1999},
 }

@inproceedings{Lia09,
 address              = {Bruges, Belgium},
 author               = {Liang, Nanying and Laurent Bougrain},
 booktitle            = {European Symposium on Artificial Neural Networks - ESANN 2009},
 pages                = {467-472},
 title                = {Decoding {F}inger {F}lexion {U}sing {A}mplitude {M}odulation From {B}and-{S}pecific {ECoG}},
 year                 = {2009},
 }

@inproceedings{Liu10,
 author               = {Liu, Chong and Hai-bin Zhao and Chun-sheng Li and Hong Wang},
 booktitle            = {Biomedical Engineering and Informatics (BMEI), 2010 3rd International Conference on},
 doi                  = {10.1109/BMEI.2010.5639943},
 keywords             = {ECoG datasets;ECoG motor imagery task classification;brain computer interface;common spatial pattern;data processing;electrocorticography;event-related desynchronization;event-related synchronization;linear support vector machine;ten-fold cross validation;training dataset;bioelectric phenomena;brain-computer interfaces;medical signal processing;support vector machines;synchronisation;},
 month                = {October},
 pages                = {804 -807},
 title                = {Classification of {ECoG} {M}otor {I}magery {T}asks {B}ased on {CSP} and {SVM}},
 volume               = {2},
 year                 = {2010},
 }

@inproceedings{Lix11,
 author               = {Li, Lijun and Dongsheng Xiong and Xiaoming Wu},
 booktitle            = {Bioinformatics and Biomedical Engineering, (iCBBE) 2011 5th International Conference on},
 doi                  = {10.1109/icbbe.2011.5780688},
 issn                 = {2151-7614},
 keywords             = {CSP algorithm;ECoG imaginary movements classification;Power Spectral Density;classification accuracy;common spatial pattern algorithm;electrocorticogram;feature extraction;human brain computer interface;motor imagery;support vector machines;brain-computer interfaces;feature extraction;medical signal processing;neurophysiology;support vector machines;},
 month                = {May},
 pages                = {1 -3},
 title                = {Classification of {I}maginary {M}ovements in {ECoG}},
 year                 = {2011},
 }

@article{Lot11,
 author               = {Lotte, F. and Cuntai Guan},
 doi                  = {10.1109/TBME.2010.2082539},
 issn                 = {0018-9294},
 journal              = {Biomedical Engineering, IEEE Transactions on},
 keywords             = {BCI competition datasets;BCI designs;Tikhonov regularization;brain-computer interfaces;common spatial patterns;electroencephalography data;feature extraction algorithms;neurophysiologically relevant spatial filters;regularized CSP;subject-to-subject transfer;brain-computer interfaces;electroencephalography;feature extraction;medical signal processing;spatial filters;Algorithms;Electroencephalography;Humans;Man-Machine Systems;Models, Neurological;Pattern Recognition, Automated;Regression Analysis;Signal Processing, Computer-Assisted;User-Computer Interface;},
 month                = {February},
 number               = {2},
 pages                = {355 -362},
 title                = {Regularizing {C}ommon {S}patial {P}atterns To {I}mprove {BCI} {D}esigns: {U}nified {T}heory and {N}ew {A}lgorithms},
 volume               = {58},
 year                 = {2011},
 }

@article{Lue06,
 author               = {Leuthardt, E.C. and K.J. Miller and G. Schalk and R.P.N. Rao and J.G. Ojemann},
 doi                  = {10.1109/TNSRE.2006.875536},
 issn                 = {1534-4320},
 journal              = {Neural Systems and Rehabilitation Engineering, IEEE Transactions on},
 keywords             = {Brain computer interfaces;Computer interfaces;Electrodes;Electroencephalography;Electromyography;Epilepsy;Spatial resolution;Speech;Stability;Systems engineering and theory;bioelectric phenomena;biomechanics;brain;handicapped aids;medical computing;medical control systems;prosthetics;ECoG;electrocorticography-based brain computer interface;motor tasks;neuroprosthetics;one-dimensional cursor movement;online control;online feedback;speech tasks;Brain amp;#8211;computer interface (BCI);brain amp;#8211;machine interface (BMI);neuroprosthetics;Cerebral Cortex;Electroencephalography;Epilepsy;Evoked Potentials;Humans;Therapy, Computer-Assisted;User-Computer Interface;Washington;},
 month                = {june},
 number               = {2},
 pages                = {194 -198},
 title                = {Electrocorticography {B}ased {B}rain {C}omputer {I}nterface-the {S}eattle {E}xperience},
 volume               = {14},
 year                 = {2006},
 }

@inproceedings{McM95,
 author               = {McMillan, G.R. and GL Calhoun and MS Middendorf and JH Schnurer and DF Ingle and VT Nasman},
 booktitle            = {Proc. RESNA 95 Annual Conf.(Vancouver, BC},
 pages                = {693--5},
 title                = {Direct {B}rain {I}nterface {U}tilizing {S}elf-{R}egulation of {S}teady-{S}tate {V}isual {E}voked {R}esponse ({SSVER})},
 year                 = {1995},
 }

@inproceedings{Men09,
 author               = {Meng, Jianjun and Guangquan Liu and Gan Huang and Xiangyang Zhu},
 booktitle            = {Robotics and Biomimetics (ROBIO), 2009 IEEE International Conference on},
 month                = {December},
 pages                = {2290 -2294},
 title                = {Automated {S}electing {S}ubset of {C}hannels {B}ased on {CSP} in {M}otor {I}magery {B}rain-{C}omputer {I}nterface {S}ystem},
 year                 = {2009},
 }

@misc{Mil08,
 accessed             = {03-June-2013},
 author               = {Miller, Kai J. and G. Schalk},
 title                = {Prediction of {F}inger {F}lexion 4th {B}rain-{C}omputer {I}nterface {D}ata {C}ompetition},
 url                  = {},
 year                 = {2008},
 }

@inproceedings{Mog06,
 address              = {New York, NY, USA},
 author               = {Moghaddam, Baback and Yair Weiss and Shai Avidan},
 booktitle            = {Proceedings of the 23rd international conference on Machine learning},
 doi                  = {http://doi.acm.org/10.1145/1143844.1143925},
 isbn                 = {1-59593-383-2},
 location             = {Pittsburgh, Pennsylvania},
 numpages             = {8},
 pages                = {641--648},
 publisher            = {ACM},
 series               = {ICML '06},
 title                = {Generalized {S}pectral {B}ounds for {S}parse {LDA}},
 year                 = {2006},
 }

@inproceedings{Mol08,
 author               = {Mollazadeh, Mohsen and Vikram Aggarwal and Girish Singhal and Andrew Law and Adam Davidson and Marc Schieber and Nitish Thakor},
 booktitle            = {Engineering in Medicine and Biology Society, 2008. EMBS 2008. 30th Annual International Conference of the IEEE},
 doi                  = {10.1109/IEMBS.2008.4650414},
 issn                 = {1557-170X},
 keywords             = {Animals;Electroencephalography;Evoked Potentials, Motor;Fingers;Macaca mulatta;Male;Motor Cortex;Motor Skills;Movement;Task Performance and Analysis;},
 month                = {August},
 pages                = {5314 -5317},
 title                = {Spectral {M}odulation of {LFP} {A}ctivity in {M1} {D}uring {D}exterous {F}inger {M}ovements},
 year                 = {2008},
 }

@article{Mor08,
 author               = {Moritz, Chet T. and Steve I. Perlmutter and Eberhard E. Fetz},
 issue                = {7222},
 journal              = {Nature},
 pages                = {639-642},
 publisher            = {Macmillan Publishers Limited. All rights reserved},
 title                = {Direct {C}ontrol of {P}aralysed {M}uscles by {C}ortical {N}eurons},
 volume               = {456},
 year                 = {2008},
 }

@misc{Mur98,
 accessed             = {03-June-2013},
 author               = {Murphy, Kevin},
 title                = {Hidden {M}arkov {M}odel ({HMM}) {T}oolbox for},
 url                  = {},
 year                 = {1998},
 }

@article{Neu01,
 accessed             = {03-June-2013},
 author               = {Neuper, C and G Pfurtscheller},
 doi                  = {10.1016/S0167-8760(01)00178-7},
 issn                 = {0167-8760},
 journal              = {International Journal of Psychophysiology},
 keywords             = {Cortical networks},
 number               = {1},
 pages                = {41 - 58},
 title                = {Event-{R}elated {D}ynamics of {C}ortical {R}hythms: {F}requency-{S}pecific {F}eatures and {F}unctional {C}orrelates},
 url                  = {},
 volume               = {43},
 year                 = {2001},
 }

@article{Nic01,
 author               = {Nicolelis, M.A.L.},
 journal              = {Nature},
 number               = {6818},
 pages                = {403--408},
 publisher            = {[London: Macmillan Journals], 1869-},
 title                = {Actions From {T}houghts},
 volume               = {409},
 year                 = {2001},
 }

@inproceedings{Ona11,
 author               = {Onaran, {\.I}brahim and N. F{\i}rat {\.I}nce and Enis \c{C}etin},
 booktitle            = {International IEEE EMBS Conference},
 month                = {August},
 title                = {Classification of {M}ultichannel {ECoG} {R}elated To {I}ndividual {F}inger {M}ovements with {R}edundant {S}patial {P}rojections},
 year                 = {2011},
 }

@article{Ona12,
 author               = {Onaran, {\.I}brahim and N. Firat Ince and A. Enis \c{C}etin},
 doi                  = {10.1016/j.bspc.2012.10.003},
 issn                 = {1746-8094},
 journal              = {Biomedical Signal Processing and Control},
 keywords             = {Unconstrained optimization},
 number               = {0},
 pages                = {-},
 title                = {Sparse {S}patial {F}ilter via a {N}ovel {O}bjective {F}unction {M}inimization with {S}mooth $\ell_1$ {R}egularization},
 year                 = {2012},
 }

@article{Pan11,
 adsnote              = {Provided by the SAO/NASA Astrophysics Data System},
 adsurl               = {http://adsabs.harvard.edu/abs/2011arXiv1105.1185P},
 archiveprefix        = {arXiv},
 author               = {Panju, M.},
 eprint               = {1105.1185},
 journal              = {ArXiv e-prints},
 keywords             = {Mathematics - Numerical Analysis},
 month                = {may},
 owner                = {onara001},
 primaryclass         = {math.NA},
 timestamp            = {2013.01.05},
 title                = {{Iterative {M}ethods for {C}omputing {E}igenvalues and {E}igenvectors}},
 year                 = {2011},
 }

@article{Pen98,
 author               = {Penny, W.D. and S.J. Roberts and MJ Stokes},
 journal              = {Journal of Neuroscience Methods},
 title                = {Imagined {H}and {M}ovements {I}dentified From the {EEG} {M}u-{R}hythm},
 year                 = {1998},
 }

@article{Pfu01,
 author               = {Pfurtscheller, G. and C. Neuper},
 doi                  = {10.1109/5.939829},
 issn                 = {0018-9219},
 journal              = {Proceedings of the IEEE},
 keywords             = {EEG classification;EEG-based brain-computer interfaces;HMM;adaptive autoregressive parameters;array of electrodes;band power;bipolar EEG recordings;direct brain-computer communication;event-related desynchronization;hand orthosis control;learning;left-hand movement;linear discrimination analysis;mental imagination;motor imagery;neural networks;neuronal activity;online feedback;parameter estimation;primary sensorimotor areas;rapid prototyping;right-hand movement;specific motor commands;tetraplegic patient;transient EEG changes;autoregressive processes;biocontrol;electroencephalography;feedback;handicapped aids;hidden Markov models;learning (artificial intelligence);medical signal processing;neural nets;neurophysiology;orthotics;signal classification;},
 month                = {July},
 number               = {7},
 pages                = {1123 -1134},
 title                = {Motor {I}magery and {D}irect {B}rain-{C}omputer {C}ommunication},
 volume               = {89},
 year                 = {2001},
 }

@article{Pfu94,
 accessed             = {03-June-2013},
 affiliation          = {Department of Medical Informatics, Institute of Biomedical Engineering Graz University of Technology Graz Austria},
 author               = {Pfurtscheller, Gert and Christa Neuper and Johannes Berger},
 issn                 = {0896-0267},
 issue                = {4},
 journal              = {Brain Topography},
 keyword              = {Biomedical and Life Sciences},
 note                 = {10.1007/BF01211172},
 pages                = {269-275},
 publisher            = {Springer New York},
 title                = {Source {L}ocalization {U}sing {E}ventrelated {D}esynchronization ({ERD}) within the {A}lpha {B}and},
 url                  = {},
 volume               = {6},
 year                 = {1994},
 }

@article{Pfu98,
 author               = {Pfurtscheller, G. and C. Neuper and A. Schlogl and K. Lugger},
 doi                  = {10.1109/86.712230},
 issn                 = {1063-6528},
 journal              = {Rehabilitation Engineering, IEEE Transactions on},
 keywords             = {adaptive signal processing;biomechanics;electroencephalography;handicapped aids;medical signal processing;parameter estimation;psychology;1 s;EEG signals separability;EEG-based cursor control;adaptive autoregressive parameters;amyotrophic lateral sclerosis patients;brain-computer interface;communication channel;computer screen cursor movement;impaired motor function;left motor imagery;linear discriminant analysis;online neural network classification;paralyzed person's communication aid;right motor imagery;sensory-motor areas;simple binary response;subject-specific frequency bands;thinking about moving;Biological neural networks;Brain computer interfaces;Communication channels;Communication system control;Electrodes;Electroencephalography;Error analysis;Frequency;Output feedback;Signal analysis},
 number               = {3},
 pages                = {316-325},
 title                = {Separability of {EEG} {S}ignals {R}ecorded {D}uring {R}ight and {L}eft {M}otor {I}magery {U}sing {A}daptive {A}utoregressive {P}arameters},
 volume               = {6},
 year                 = {1998},
 }

@article{Pfu99,
 accessed             = {03-June-2013},
 author               = {Pfurtscheller, G. and F.H. Lopes da Silva},
 doi                  = {10.1016/S1388-2457(99)00141-8},
 issn                 = {1388-2457},
 journal              = {Clinical Neurophysiology},
 keywords             = {Brain oscillations},
 number               = {11},
 pages                = {1842 - 1857},
 title                = {Event-{R}elated {EEG}/{MEG} {S}ynchronization and {D}esynchronization: {B}asic {P}rinciples},
 url                  = {},
 volume               = {110},
 year                 = {1999},
 }

@book{Pfu99a,
 accessed             = {03-June-2013},
 author               = {Pfurtscheller, G. and F.H.L. da Silva},
 isbn                 = {9780444829993},
 lccn                 = {99019484},
 publisher            = {Elsevier Science Health Science Division},
 series               = {Event-related Desynchronization},
 title                = {Event {R}elated {D}esynchronization},
 url                  = {},
 year                 = {1999},
 }

@article{Pis12,
 accessed             = {03-June-2013},
 author               = {Pistohl, Tobias and Andreas Schulze-Bonhage and Ad Aertsen and Carsten Mehring and Tonio Ball},
 doi                  = {10.1016/j.neuroimage.2011.06.084},
 issn                 = {1053-8119},
 journal              = {NeuroImage},
 keywords             = {BMI},
 number               = {1},
 pages                = {248 - 260},
 title                = {Decoding {N}atural {G}rasp {T}ypes from {H}uman {ECoG}},
 url                  = {},
 volume               = {59},
 year                 = {2012},
 }

@misc{Pol95,
 author               = {Polikoff, James B. and H. Timothy Bunnell and Winslow J. Borkowski Jr.},
 title                = {Toward a {P300}-{B}ased {C}omputer {I}nterface},
 year                 = {1995},
 }

@article{Rab89,
 author               = {Rabiner, L. R.},
 journal              = {Proceedings of the IEEE},
 number               = {2},
 pages                = {257-286},
 title                = {A {T}utorial on {H}idden {M}arkov {M}odels and {S}elected},
 volume               = {77},
 year                 = {1989},
 }

@article{Ram00,
 author               = {Ramoser, H. and J. M\"uller-Gerking and G. Pfurtscheller},
 booktitle            = {Rehabilitation Engineering, IEEE Transactions on [see also IEEE Trans. on Neural Systems and Rehabilitation]},
 journal              = {IEEE Transactions on Rehabilitation Engineering},
 month                = {December},
 number               = {4},
 pages                = {441--446},
 title                = {Optimal {S}patial {F}iltering of {S}ingle {T}rial {EEG} {D}uring {I}magined {H}and {M}ovement},
 volume               = {8},
 year                 = {2000},
 }

@misc{Reu08,
 address              = {Enschede},
 author               = {Reuderink, B. and M. Poel},
 journal              = {Internal Report},
 month                = {July},
 number               = {TR-CTI},
 organization         = {Univ. of Twente},
 publisher            = {Centre for Telematics and Information Technology, University of Twente},
 title                = {Robustness of the {C}ommon {S}patial {P}atterns {A}lgorithm in the {BCI}-{P}ipeline},
 year                 = {2008},
 }

@article{Rif04,
 accessed             = {03-June-2013},
 author               = {Rifkin, Ryan and Aldebaro Klautau},
 issn                 = {1532-4435},
 journal              = {J. Mach. Learn. Res.},
 month                = {December},
 numpages             = {41},
 pages                = {101--141},
 publisher            = {JMLR.org},
 title                = {In {D}efense of {O}ne-{V}s-{A}ll {C}lassification},
 url                  = {},
 volume               = {5},
 year                 = {2004},
 }

@conference{Ros13,
 author               = {Rossi, M. and A. M. Haimovich and Y. C. Eldar},
 journal              = {ICASSP},
 title                = {Conditions for {T}arget {R}ecovery in {S}patial {C}ompressive {S}ensing for {MIMO} {R}adar},
 year                 = {2013},
 }

@article{Rud92,
 author               = {Leonid I. Rudin and Stanley Osher and Emad Fatemi},
 doi                  = {10.1016/0167-2789(92)90242-F},
 issn                 = {0167-2789},
 journal              = {Physica D: Nonlinear Phenomena},
 number               = {1â€“4},
 pages                = {259 - 268},
 title                = {Nonlinear total variation based noise removal algorithms},
 url                  = {},
 volume               = {60},
 year                 = {1992},
 }

@article{San11,
 accessed             = {03-June-2013},
 author               = {C. Sannelli, C. Vidaurre, K-R Mueller, B. Blankertz},
 journal              = {Journal of Neural Engineering},
 pages                = {7pp},
 title                = {{CSP} {P}atches: an {E}nsemble of {O}ptimized {S}patial {F}ilters. An {E}valuation {S}tudy.},
 url                  = {},
 volume               = {8(2):025012},
 year                 = {2011},
 }

@article{Sch05,
 author               = {Scherberger, H. and M.R. Jarvis and R.A. Andersen},
 journal              = {Neuron},
 number               = {2},
 pages                = {347--354},
 publisher            = {Elsevier},
 title                = {Cortical {L}ocal {F}ield {P}otential {E}ncodes {M}ovement {I}ntentions in the {P}osterior {P}arietal {C}ortex},
 volume               = {46},
 year                 = {2005},
 }

@misc{Sch09,
 author               = {Schmidt, Mark and Glenn Fung and Rmer Rosales},
 title                = {Fast {O}ptimization {M}ethods for {L1} {R}egularization: A {C}omparative {S}tudy and {T}wo {N}ew {A}pproaches},
 year                 = {2009},
 }

@article{Sez82,
 author               = {I. Sezan and H. Stark},
 journal              = {IEEE Transactions on Medical Imaging},
 number               = {2},
 pages                = {95-101},
 title                = {Image Restoration by the Method of Convex Projections: Part 2-Applications and Numerical Results},
 volume               = {1},
 year                 = {1982},
 }

@article{She03,
 author               = {Shenoy, Krishna V. and Daniella Meeker and Shiyan Cao and Sohaib A. Kureshi and Bijan Pesaran and Christopher A. Buneo and Aaron P. Batista and Partha P. Mitra and Joel W. Burdick and Richard A. Andersen},
 journal              = {NeuroReport},
 number               = {4},
 pages                = {591-596},
 title                = {Neural {P}rosthetic {C}ontrol {S}ignals From {P}lan {A}ctivity},
 volume               = {14},
 year                 = {2003},
 }

@inproceedings{She07,
 author               = {Shenoy, P. and K.J. Miller and J.G. Ojemann and R.P.N. Rao},
 booktitle            = {Neural Engineering, 2007. CNE '07. 3rd International IEEE/EMBS Conference on},
 doi                  = {10.1109/CNE.2007.369644},
 keywords             = {brain computer interface;electrocorticographic signals;finger movement classification;biocontrol;electroencephalography;human computer interaction;motion compensation;},
 month                = {May},
 pages                = {192 -195},
 title                = {Finger {M}ovement {C}lassification for an {E}lectrocorticographic {BCI}},
 year                 = {2007},
 }

@article{Sla08,
 author               = {Slavakis, K. and S. Theodoridis and I. Yamada},
 issn                 = {1053-587X},
 issue                = {7},
 journal              = {IEEE Transactions on Signal Processing},
 pages                = {2781--2796},
 part                 = {1},
 title                = {Online {K}ernel-{B}ased {C}lassification {U}sing {A}daptive {P}rojection {A}lgorithms},
 volume               = {56},
 year                 = {2008},
 }

@article{Sla09,
 address              = {Piscataway, NJ, USA},
 author               = {Slavakis, Konstantinos and Theodoridis, Sergios and Yamada, Isao},
 doi                  = {10.1109/TSP.2009.2027771},
 issn                 = {1053-587X},
 journal              = {IEEE Transactions on Signal Processing},
 keywords             = {Adaptive learning, adaptive learning, beamforming, convex analysis, fixed-point set, reproducing kernel Hilbert space (RKHS)},
 month                = {dec},
 number               = {12},
 numpages             = {21},
 pages                = {4744--4764},
 publisher            = {IEEE Press},
 title                = {Adaptive constrained learning in reproducing Kernel Hilbert spaces: the robust beamforming case},
 url                  = {},
 volume               = {57},
 year                 = {2009},
 }

@inproceedings{Som00,
 author               = {Somol, P. and P. Pudil},
 booktitle            = {Pattern Recognition, 2000. Proceedings. 15th International Conference on},
 organization         = {IEEE},
 pages                = {406--409},
 title                = {Oscillating {S}earch {A}lgorithms for {F}eature {S}election},
 volume               = {2},
 year                 = {2000},
 }

@inproceedings{Som08,
 author               = {Somol, P. and J. Novovicov{\'a} and J. Grim and P. Pudil},
 booktitle            = {Pattern Recognition, 2008. ICPR 2008. 19th International Conference on},
 organization         = {IEEE},
 pages                = {1--4},
 title                = {Dynamic {O}scillating {S}earch {A}lgorithm for {F}eature {S}election},
 year                 = {2008},
 }

@article{Sut92,
 accessed             = {03-June-2013},
 author               = {Sutter, Erich E.},
 doi                  = {10.1016/0745-7138(92)90045-7},
 issn                 = {0745-7138},
 journal              = {Journal of Microcomputer Applications},
 note                 = {Special Issue on Computers for Handicapped People},
 number               = {1},
 pages                = {31 - 45},
 title                = {The {B}rain {R}esponse {I}nterface: {C}ommunication {T}hrough {V}isually-{I}nduced {E}lectrical {B}rain {R}esponses},
 url                  = {},
 volume               = {15},
 year                 = {1992},
 }

@article{The11,
 author               = {K. S Theodoridis and I. Yamada},
 journal              = {IEEE Signal Processing Magazine},
 number               = {1},
 pages                = {97-123},
 title                = {Adaptive Learning in a World of Projections},
 volume               = {28},
 year                 = {2011},
 }

@misc{Tom06,
 author               = {Tomioka, Ryota and Guido Dornhege and Guido Nolte and Benjamin Blankertz and Kazuyuki Aihara and Klaus-Robert M\"uller},
 title                = {Spectrally {W}eighted {C}ommon {S}patial {P}attern {A}lgorithm for {S}ingle {T}rial {EEG} {C}lassification},
 year                 = {2006},
 }

@article{Tow11,
 author               = {Townsend, B.R. and E. Subasi and H. Scherberger},
 journal              = {The Journal of Neuroscience},
 number               = {40},
 pages                = {14386--14398},
 publisher            = {Society for Neuroscience},
 title                = {Grasp {M}ovement {D}ecoding from {P}remotor and {P}arietal {C}ortex},
 volume               = {31},
 year                 = {2011},
 }

@article{Tru85,
 author               = {Trussell, H.J. and M. R. Civanlar},
 journal              = {IEEE Transactions on Acoustics, Speech and Signal Processing},
 number               = {6},
 pages                = {1632-1634},
 title                = {The {L}andweber {I}teration and {P}rojection {O}nto {C}onvex {S}et},
 volume               = {33},
 year                 = {1985},
 }

@article{Tuy81,
 author               = {A. Lent and H. Tuy},
 journal              = {Journal of Mathematical Analysis and Applications, 83 (2), pp.1981},
 pages                = {554--565},
 title                = {An {I}terative {M}ethod for the {E}xtrapolation of {B}and-{L}imited {F}unctions},
 volume               = {83},
 year                 = {1981},
 }

@article{Vel08,
 author               = {Velliste, M. and S. Perel and M.C. Spalding and A.S. Whitford and A.B. Schwartz},
 journal              = {Nature},
 number               = {7198},
 pages                = {1098--1101},
 publisher            = {Nature Publishing Group},
 title                = {Cortical {C}ontrol of a {P}rosthetic {A}rm for {S}elf-{F}eeding},
 volume               = {453},
 year                 = {2008},
 }

@inproceedings{Wan05,
 author               = {Wang, Yijun and Shangkai Gao and Xiaorong Gao},
 booktitle            = {Engineering in Medicine and Biology Society, 2005. IEEE-EMBS 2005. 27th Annual International Conference of the},
 doi                  = {10.1109/IEMBS.2005.1615701},
 keywords             = {EEG;brain-computer interface;channel reduction;channel selection;common spatial pattern method;electroencephalogram;event-related desynchronization;imagination tasks;imagined foot movements;imagined hand movements;linear discriminant analysis;motor imagery;readiness potential;signal classification;bioelectric potentials;electroencephalography;handicapped aids;medical signal processing;signal classification;},
 month                = {January},
 pages                = {5392 -5395},
 title                = {Common {S}patial {P}attern {M}ethod for {C}hannel {S}election in {M}otor {I}magery {B}ased {B}rain-{C}omputer {I}nterface},
 year                 = {2005},
 }

@inproceedings{Wan09,
 author               = {Wang, W. and A.D. Degenhart and J.L. Collinger and R. Vinjamuri and G.P. Sudre and P.D. Adelson and D.L. Holder and E.C. Leuthardt and D.W. Moran and M.L. Boninger and A.B. Schwartz and D.J. Crammond and E.C. Tyler-Kabara and D.J. Weber},
 booktitle            = {Engineering in Medicine and Biology Society, 2009. EMBC 2009. Annual International Conference of the IEEE},
 issn                 = {1557-170X},
 month                = {September},
 pages                = {586 -589},
 title                = {Human {M}otor {C}ortical {A}ctivity {R}ecorded with {M}icro-{ECoG} {E}lectrodes, {D}uring {I}ndividual {F}inger {M}ovements},
 year                 = {2009},
 }

@article{Wan99,
 author               = {Wang, Yunhua and Patrick Berg and Michael Scherg},
 issn                 = {1388-2457},
 journal              = {Clinical Neurophysiology},
 keywords             = {Covariance matrix analysis},
 number               = {4},
 pages                = {604 - 614},
 title                = {Common {S}patial {S}ubspace {D}ecomposition {A}pplied To {A}nalysis of {B}rain {R}esponses {U}nder {M}ultiple {T}ask {C}onditions: a {S}imulation {S}tudy},
 volume               = {110},
 year                 = {1999},
 }

@article{Wei04,
 author               = {Weiskopf, N. and K. Mathiak and S.W. Bock and F. Scharnowski and R. Veit and W. Grodd and R. Goebel and N. Birbaumer},
 doi                  = {10.1109/TBME.2004.827063},
 issn                 = {0018-9294},
 journal              = {Biomedical Engineering, IEEE Transactions on},
 keywords             = {Biomedical imaging;Blood;Brain computer interfaces;Delay;Electroencephalography;Hemodynamics;Magnetic resonance imaging;Neurofeedback;Psychology;Spatial resolution;bioelectric potentials;biomedical MRI;brain;feedback;haemodynamics;handicapped aids;medical image processing;neurophysiology;brain-computer interface;data processing;differential feedback;hemodynamic brain activity;neurofeedback;parahippocampal place area;real-time functional magnetic resonance imaging;self-regulation;supplementary motor area;Adult;Biofeedback (Psychology);Brain;Brain Mapping;Feasibility Studies;Feedback;Female;Hippocampus;Humans;Image Interpretation, Computer-Assisted;Magnetic Resonance Imaging;Male;Motor Cortex;Online Systems;Reproducibility of Results;Sensitivity and Specificity;User-Computer Interface;},
 month                = {june},
 number               = {6},
 pages                = {966 -970},
 title                = {Principles of a {B}rain-{C}omputer {I}nterface ({BCI}) {B}ased on {R}eal-{T}ime {F}unctional {M}agnetic {R}esonance {I}maging ({fMRI})},
 volume               = {51},
 year                 = {2004},
 }

@article{Wes02,
 author               = {Wessberg, J. and C.R. Stambaugh and J.D. Kralik and P.D. Beck and M. Laubach and J.K. Chapin and J. Kim and S.J. Biggs and M.A. Srinivasan and M.A.L. Nicolelis},
 journal              = {Nature},
 pages                = {361--365},
 title                = {Letters To {N}ature: {R}eal-{T}ime {P}rediction of {H}and {T}rajectory by {E}nsembles of {C}ortical {N}eurons in {P}rimates},
 year                 = {2002},
 }

@article{Wol00,
 author               = {Wolpaw, Jonathan R. and Niels Birbaumer and William J. Heetderks and Dennis J. McFarland and P. Hunter Peckham and Gerwin Schalk and Emanuel Donchin and Louis A. Quatrano and Charles J. Robinson and Theresa M. Vaughan},
 journal              = {IEEE TRANSACTIONS ON REHABILITATION ENGINEERING},
 month                = {June},
 number               = {2},
 title                = {Brain-{C}omputer {I}nterface {T}echnology: A {R}eview of the {F}irst {I}nternational {M}eeting},
 volume               = {8},
 year                 = {2000},
 }

@article{Wol94,
 author               = {Wolpaw, Jonathan R. and Dennis J. McFarland},
 doi                  = {10.1016/0013-4694(94)90135-X},
 issn                 = {0013-4694},
 journal              = {Electroencephalography and Clinical Neurophysiology},
 keywords             = {Electroencephalography},
 number               = {6},
 pages                = {444 - 449},
 title                = {Multichannel {EEG}-{B}ased {B}rain-{C}omputer {C}ommunication},
 volume               = {90},
 year                 = {1994},
 }

@article{Yan11,
 author               = {Yanagisawa, T. and M. Hirata and Y. Saitoh and T. Goto and H. Kishima and R. Fukuma and H. Yokoi and Y. Kamitani and T. Yoshimine},
 journal              = {Journal of neurosurgery},
 number               = {6},
 pages                = {1715--1722},
 publisher            = {American Association of Neurological Surgeons},
 title                = {Real-{T}ime {C}ontrol of a {P}rosthetic {H}and {U}sing {H}uman {E}lectrocorticography {S}ignals},
 volume               = {114},
 year                 = {2011},
 }

@article{Yin08,
 author               = {Yin, W. and Osher, S. and Goldfarb, D. and Darbon, J.},
 doi                  = {10.1137/070703983},
 eprint               = {http://epubs.siam.org/doi/pdf/10.1137/070703983},
 journal              = {SIAM Journal on Imaging Sciences},
 number               = {1},
 pages                = {143-168},
 title                = {Bregman Iterative Algorithms for $\ell_1$-Minimization with Applications to Compressed Sensing},
 url                  = {},
 volume               = {1},
 year                 = {2008},
 }

@inproceedings{Yon08,
 author               = {Yong, Xinyi and R.K. Ward and G.E. Birch},
 booktitle            = {Acoustics, Speech and Signal Processing, 2008. ICASSP 2008. IEEE International Conference on},
 month                = {April},
 pages                = {417 -420},
 title                = {Sparse {S}patial {F}ilter {O}ptimization for {EEG} {C}hannel {R}eduction in {B}rain-{C}omputer {I}nterface},
 year                 = {2008},
 }

@inproceedings{You11,
 author               = {Shin, Younghak and Seungchan Lee and Minkyu Ahn and Sung Chan Jun and Heung-No Lee},
 booktitle            = {Noninvasive Functional Source Imaging of the Brain and Heart 2011 8th International Conference on Bioelectromagnetism (NFSI ICBEM), 2011 8th International Symposium on},
 doi                  = {10.1109/NFSI.2011.5936827},
 keywords             = {EEG signals;beta rhythm;brain-computer interface;classification methods;communication channel;control channel;electroencephalogram;linear discriminant analysis;minimization;motor imagery based BCI classification;sensorimotor rhythm analysis;sparse representation;brain-computer interfaces;electroencephalography;medical signal processing;minimisation;signal classification;},
 month                = {May},
 pages                = {93 -97},
 title                = {Motor {I}magery {B}ased {BCI} {C}lassification Via {S}parse {R}epresentation of {EEG} {S}ignals},
 year                 = {2011},
 }

@article{You82,
 author               = {Youla, D.C. and H. Webb},
 doi                  = {10.1109/TMI.1982.4307555},
 issn                 = {0278-0062},
 journal              = {Medical Imaging, IEEE Transactions on},
 keywords             = {Computer science;Convergence of numerical methods;Degradation;Hilbert space;Image reconstruction;Image restoration;Iterative algorithms;Microwave imaging;Sufficient conditions;Vectors},
 number               = {2},
 pages                = {81-94},
 title                = {Image {R}estoration by the {M}ethod of {C}onvex {P}rojections: {P}art 1 {N}um2014;Theory},
 volume               = {1},
 year                 = {1982},
 }

@inproceedings{Yua05,
 author               = {Li, Yuanqing and Cuntai Guan and Jianzhao Qin},
 booktitle            = {Engineering in Medicine and Biology Society, 2005. IEEE-EMBS 2005. 27th Annual International Conference of the},
 doi                  = {10.1109/IEMBS.2005.1615686},
 keywords             = {EEG;brain-computer interface;cursor control;dynamical common spatial pattern feature;dynamical power feature;feature extraction;feature vector;high classification accuracy;signal preprocessing;sparse component analysis;electroencephalography;feature extraction;handicapped aids;medical signal processing;signal classification;},
 month                = {January},
 pages                = {5335 -5338},
 title                = {Enhancing {F}eature {E}xtraction with {S}parse {C}omponent {A}nalysis for {B}rain-{C}omputer {I}nterface},
 year                 = {2005},
 }

 @article{censor1987optimization,
  title={Optimization of “$\backslash$logx” Entropy over Linear Equality Constraints},
  author={Censor, Yair and Lent, Arnold},
  journal={SIAM Journal on Control and Optimization},
  volume={25},
  number={4},
  pages={921--933},
  year={1987},
  publisher={SIAM}
}
@article{censor1987some,
  title={On some optimization techniques in image reconstruction from projections},
  author={Censor, Yair and Herman, Gabor T},
  journal={Applied Numerical Mathematics},
  volume={3},
  number={5},
  pages={365--391},
  year={1987},
  publisher={Elsevier}
}

@article{censor1992proximal,
  title={Proximal minimization algorithm withD-functions},
  author={Censor, Yair and Zenios, Stavros Andrea},
  journal={Journal of Optimization Theory and Applications},
  volume={73},
  number={3},
  pages={451--464},
  year={1992},
  publisher={Springer}
}

@article{censor1981row,
  title={Row-action methods for huge and sparse systems and their applications},
  author={Censor, Yair},
  journal={SIAM review},
  volume={23},
  number={4},
  pages={444--466},
  year={1981},
  publisher={SIAM}
}

@article{censor1991optimization,
  title={Optimization of Burg's entropy over linear constraints},
  author={Censor, Yair and De Pierro, Alvaro R and Iusem, Alfredo N},
  journal={Applied Numerical Mathematics},
  volume={7},
  number={2},
  pages={151--165},
  year={1991},
  publisher={Elsevier}
}
@incollection{yamada2011minimizing,
  title={Minimizing the Moreau envelope of nonsmooth convex functions over the fixed point set of certain quasi-nonexpansive mappings},
  author={Yamada, Isao and Yukawa, Masahiro and Yamagishi, Masao},
  booktitle={Fixed-Point Algorithms for Inverse Problems in Science and Engineering},
  pages={345--390},
  year={2011},
  publisher={Springer}
}


@article{Chambolle,
 author = {Chambolle, Antonin},
 title = {An Algorithm for Total Variation Minimization and Applications},
 journal = {J. Math. Imaging Vis.},
 issue_date = {January-March 2004},
 volume = {20},
 number = {1-2},
 month = jan,
 year = {2004},
 issn = {0924-9907},
 pages = {89--97},
 numpages = {9},
 url = {},
 doi = {10.1023/B:JMIV.0000011325.36760.1e},
 acmid = {964985},
 publisher = {Kluwer Academic Publishers},
 address = {Norwell, MA, USA},
 keywords = {denoising, image reconstruction, mean curvature motion, total variation, zooming},
}


\begin{thebibliography}{10}
\providecommand{\url}[1]{#1}
\csname url@samestyle\endcsname
\providecommand{\newblock}{\relax}
\providecommand{\bibinfo}[2]{#2}
\providecommand{\BIBentrySTDinterwordspacing}{\spaceskip=0pt\relax}
\providecommand{\BIBentryALTinterwordstretchfactor}{4}
\providecommand{\BIBentryALTinterwordspacing}{\spaceskip=\fontdimen2\font plus
\BIBentryALTinterwordstretchfactor\fontdimen3\font minus
  \fontdimen4\font\relax}
\providecommand{\BIBforeignlanguage}[2]{{%
\expandafter\ifx\csname l@#1\endcsname\relax
\typeout{** WARNING: IEEEtran.bst: No hyphenation pattern has been}%
\typeout{** loaded for the language `#1'. Using the pattern for}%
\typeout{** the default language instead.}%
\else
\language=\csname l@#1\endcsname
\fi
#2}}
\providecommand{\BIBdecl}{\relax}
\BIBdecl

\bibitem{Rud92}
L.~I. Rudin, S.~Osher, and E.~Fatemi, ``Nonlinear total variation based noise
  removal algorithms,'' \emph{Physica D: Nonlinear Phenomena}, vol.~60, no.
  1–4, pp. 259 -- 268, 1992.

\bibitem{Bar07}
R.~Baraniuk, ``Compressive sensing [lecture notes],'' \emph{Signal Processing
  Magazine, IEEE}, vol.~24, no.~4, pp. 118--121, 2007.

\bibitem{Can08}
E.~Candes and M.~Wakin, ``An introduction to compressive sampling,''
  \emph{Signal Processing Magazine, IEEE}, vol.~25, no.~2, pp. 21--30, 2008.

\bibitem{Kos12}
K.~Kose, V.~Cevher, and A.~Cetin, ``Filtered variation method for denoising and
  sparse signal processing,'' \emph{Acoustics, Speech and Signal Processing
  (ICASSP), 2012 IEEE International Conference on}, pp. 3329--3332, 2012.

\bibitem{Gunay}
O.~G\"{u}nay, K.~K\"{o}se, B.~U. T\"{o}reyin, and A.~E. \c{C}etin,
  ``Entropy-functional-based online adaptive decision fusion framework with
  application to wildfire detection in video,'' \emph{IEEE Transactions on
  Image Processing}, vol.~21, no.~5, pp. 2853--2865, May 2012.

\bibitem{Bre67}
L.~Bregman, ``The {R}elaxation {M}ethod of {F}inding the {C}ommon {P}oint of
  {C}onvex {S}ets and {I}ts {A}pplication to the {S}olution of {P}roblems in
  {C}onvex {P}rogramming,'' \emph{\{USSR\} Computational Mathematics and
  Mathematical Physics}, vol.~7, no.~3, pp. 200 -- 217, 1967.

\bibitem{Yin08}
W.~Yin, S.~Osher, D.~Goldfarb, and J.~Darbon, ``Bregman iterative algorithms
  for $\ell_1$-minimization with applications to compressed sensing,''
  \emph{SIAM Journal on Imaging Sciences}, vol.~1, no.~1, pp. 143--168, 2008.

\bibitem{Kiv12}
K.~K\"{o}se, ``Signal and {I}mage {P}rocessing {A}lgorithms {U}sing {I}nterval
  {C}onvex {P}rogramming and {S}parsity,'' Ph.D. dissertation, Bilkent
  University, 2012.

\bibitem{Bregman}
L.~Bregman, ``Finding the common point of convex sets by the method of
  successive projection.(russian),'' \emph{\{USSR\} Dokl. Akad. Nauk SSSR},
  vol.~7, no.~3, pp. 200 -- 217, 1965.

\bibitem{You82}
D.~Youla and H.~Webb, ``Image {R}estoration by the {M}ethod of {C}onvex
  {P}rojections: {P}art 1 {N}um2014;theory,'' \emph{Medical Imaging, IEEE
  Transactions on}, vol.~1, no.~2, pp. 81--94, 1982.

\bibitem{Her95}
G.~T. Herman, ``Image {R}econstruction from {P}rojections,'' \emph{Real-Time
  Imaging}, vol.~1, no.~1, pp. 3--18, 1995.

\bibitem{Cen12}
Y.~Censor, W.~Chen, P.~L. Combettes, R.~Davidi, and G.~Herman,
  ``\BIBforeignlanguage{English}{On the {E}ffectiveness of {P}rojection
  {M}ethods for {C}onvex {F}easibility {P}roblems with {L}inear {I}nequality
  {C}onstraints},'' \emph{\BIBforeignlanguage{English}{Computational
  Optimization and Applications}}, vol.~51, no.~3, pp. 1065--1088, 2012.

\bibitem{Sla08}
K.~Slavakis, S.~Theodoridis, and I.~Yamada, ``Online {K}ernel-{B}ased
  {C}lassification {U}sing {A}daptive {P}rojection {A}lgorithms,'' \emph{IEEE
  Transactions on Signal Processing}, vol.~56, pp. 2781--2796, 2008.

\bibitem{Cet03}
A.~\c{C}etin, H.~\"Ozakta\c{s}, and H.~Ozaktas, ``Resolution {E}nhancement of
  {L}ow {R}esolution {W}avefields with,'' \emph{Electronics Letters}, vol.~39,
  no.~25, pp. 1808--1810, 2003.

\bibitem{Cetin94}
A.~E. Cetin and R.~Ansari, ``Signal recovery from wavelet transform maxima,''
  \emph{IEEE Transactions on Signal Processing}, pp. 673--676, 1994.

\bibitem{Cetin89}
A.~E. Cetin, ``Reconstruction of signals from fourier transform samples,''
  \emph{Signal Processing}, pp. 129--148, 1989.

\bibitem{Kose11}
K.~Kose and A.~E. Cetin, ``Low-pass filtering of irregularly sampled signals
  using a set theoretic framework,'' \emph{IEEE Signal Processing Magazine},
  pp. 117--121, 2011.

\bibitem{Cen81}
Y.~Censor and A.~Lent, ``An {I}terative {R}ow-{A}ction {M}ethod for {I}nterval
  {C}onvex {P}rogramming,'' \emph{Journal of Optimization Theory and
  Applications}, vol.~34, no.~3, pp. 321--353, 1981.

\bibitem{Sla09}
K.~Slavakis, S.~Theodoridis, and I.~Yamada, ``Adaptive constrained learning in
  reproducing kernel hilbert spaces: the robust beamforming case,'' \emph{IEEE
  Transactions on Signal Processing}, vol.~57, no.~12, pp. 4744--4764, dec
  2009.

\bibitem{The11}
K.~S. Theodoridis and I.~Yamada, ``Adaptive learning in a world of
  projections,'' \emph{IEEE Signal Processing Magazine}, vol.~28, no.~1, pp.
  97--123, 2011.

\bibitem{censor1987optimization}
Y.~Censor and A.~Lent, ``Optimization of "$\backslash$logx" entropy over linear
  equality constraints,'' \emph{SIAM Journal on Control and Optimization},
  vol.~25, no.~4, pp. 921--933, 1987.

\bibitem{Tru85}
H.~Trussell and M.~R. Civanlar, ``The {L}andweber {I}teration and {P}rojection
  {O}nto {C}onvex {S}et,'' \emph{IEEE Transactions on Acoustics, Speech and
  Signal Processing}, vol.~33, no.~6, pp. 1632--1634, 1985.

\bibitem{Com04}
P.~L. Combettes and J.~Pesquet, ``Image restoration subject to a total
  variation constraint,'' \emph{IEEE Transactions on Image Processing},
  vol.~13, pp. 1213--1222, 2004.

\bibitem{Com93}
P.~Combettes, ``The foundations of set theoretic estimation,''
  \emph{Proceedings of the IEEE}, vol.~81, no.~2, pp. 182 --208, February 1993.

\bibitem{Kim92}
A.~E. Cetin and R.~Ansari, ``Convolution based framework for signal recovery
  and applications,'' \emph{JOSA-A}, pp. 673--676, 1988.

\bibitem{yamada2011minimizing}
I.~Yamada, M.~Yukawa, and M.~Yamagishi, ``Minimizing the moreau envelope of
  nonsmooth convex functions over the fixed point set of certain
  quasi-nonexpansive mappings,'' in \emph{Fixed-Point Algorithms for Inverse
  Problems in Science and Engineering}.\hskip 1em plus 0.5em minus 0.4em\relax
  Springer, 2011, pp. 345--390.

\bibitem{censor1987some}
Y.~Censor and G.~T. Herman, ``On some optimization techniques in image
  reconstruction from projections,'' \emph{Applied Numerical Mathematics},
  vol.~3, no.~5, pp. 365--391, 1987.

\bibitem{Sez82}
I.~Sezan and H.~Stark, ``Image restoration by the method of convex projections:
  Part 2-applications and numerical results,'' \emph{IEEE Transactions on
  Medical Imaging}, vol.~1, no.~2, pp. 95--101, 1982.

\bibitem{censor1992proximal}
Y.~Censor and S.~A. Zenios, ``Proximal minimization algorithm
  withd-functions,'' \emph{Journal of Optimization Theory and Applications},
  vol.~73, no.~3, pp. 451--464, 1992.

\bibitem{Tuy81}
A.~Lent and H.~Tuy, ``An {I}terative {M}ethod for the {E}xtrapolation of
  {B}and-{L}imited {F}unctions,'' \emph{Journal of Mathematical Analysis and
  Applications, 83 (2), pp.1981}, vol.~83, pp. 554--565, 1981.

\bibitem{censor1981row}
Y.~Censor, ``Row-action methods for huge and sparse systems and their
  applications,'' \emph{SIAM review}, vol.~23, no.~4, pp. 444--466, 1981.

\bibitem{censor1991optimization}
Y.~Censor, A.~R. De~Pierro, and A.~N. Iusem, ``Optimization of burg's entropy
  over linear constraints,'' \emph{Applied Numerical Mathematics}, vol.~7,
  no.~2, pp. 151--165, 1991.

\bibitem{Ros13}
M.~Rossi, A.~M. Haimovich, and Y.~C. Eldar, ``Conditions for {T}arget
  {R}ecovery in {S}patial {C}ompressive {S}ensing for {MIMO} {R}adar,'' 2013.

\bibitem{Gub67}
L.~Gubin, B.~Polyak, and E.~Raik, ``The {M}ethod of {P}rojections for {F}inding
  the {C}ommon {P}oint of {C}onvex {S}ets,'' \emph{\{USSR\} Computational
  Mathematics and Mathematical Physics}, vol.~7, no.~6, pp. 1 -- 24, 1967.

\bibitem{Com12}
P.~L. Combettes, ``Algorithmes proximaux pour l\'{e}s problemes d\'{
  }optimisation structur les,'' 2012.

\bibitem{Cet97}
A.~E. \c{C}etin, O.~Gerek, and Y.~Yardimci, ``Equiripple {FIR} {F}ilter
  {D}esign by the {FFT} {A}lgorithm,'' \emph{IEEE Signal Processing Magazine},
  vol.~14, no.~2, pp. 60--64, 1997.

\bibitem{Chambolle}
A.~Chambolle, ``An algorithm for total variation minimization and
  applications,'' \emph{J. Math. Imaging Vis.}, vol.~20, no. 1-2, pp. 89--97,
  Jan. 2004.

\bibitem{Com11}
P.~L. Combettes and J.-C. Pesquet, ``\BIBforeignlanguage{English}{Proximal
  splitting methods in signal processing},'' in
  \emph{\BIBforeignlanguage{English}{Fixed-Point Algorithms for Inverse
  Problems in Science and Engineering}}, ser. Springer Optimization and Its
  Applications, H.~H. Bauschke, R.~S. Burachik, P.~L. Combettes, V.~Elser,
  D.~R. Luke, and H.~Wolkowicz, Eds.\hskip 1em plus 0.5em minus 0.4em\relax
  Springer New York, 2011, pp. 185--212.

\end{thebibliography}

\end{document}